\documentclass[alpha-refs]{wiley-article}

\usepackage{color}
\usepackage{ulem}
\usepackage{amsmath,esint}
\usepackage{comment}
\usepackage{xr}

\usepackage{siunitx}

\papertype{Original Article}
\paperfield{Journal Section}

\title{Hamilton's Principle with Phase Changes and 
Conservation Principles for Moist Potential Vorticity}

\author[1\authfn{1}]{Parvathi Kooloth}
\author[1,2]{Leslie M. Smith}
\author[1,3]{Samuel N. Stechmann}

\affil[1]{Department of Mathematics, University of Wisconsin--Madison, Madison, WI, 53706, USA}
\affil[2]{Department of Engineering Physics, University of Wisconsin--Madison, Madison, WI, 53706, USA}
\affil[3]{Department of Atmospheric and Oceanic Sciences, University of Wisconsin--Madison, Madison, WI, 53706, USA}

\corraddress{Parvathi Kooloth, Department of Mathematics, University of Wisconsin Madison, Madison, WI, 53706, USA}
\corremail{parvathi.kooloth@wisc.edu}

\fundinginfo{US National Science Foundation, Division of Mathematical Sciences, Grant/Award Number: DMS-1907667
\\[12pt]
Submitted on March 12, 2022}

\runningauthor{Kooloth et al.}

\begin{document}

\maketitle

\begin{abstract}
Many definitions of moist potential vorticity (PV) have been proposed
to extend the dry theory of 
Ertel PV. None of the moist PV definitions seem to have all of the desirable properties of the dry Ertel PV.
For instance, dry PV is not only a globally conserved quantity, but also a material invariant that is conserved along fluid parcel trajectories. 
Therefore, an open question remains:
is there a 
moist PV that is a material invariant,
if clouds and phase changes of water are present?
In prior studies, definitions of moist PV have been proposed 
based on physical and mathematical intuition.
Here, a systematic approach is used. In particular, a
particle relabeling symmetry is devised for a moist atmosphere 
and then Noether's theorem is
employed to arrive at the associated conservation laws 
for moist PV. A priori, it is not clear whether this systematic approach will be viable, since it relies on variational derivatives in Hamilton's principle, and  
phase changes introduce singularities 
that could potentially prevent derivatives 
at cloud edge.
However, it is shown that the energy and the Lagrangian density
are sufficiently smooth to allow variational derivatives,
in a moist Boussinesq system with reversible phase transitions
between water vapor and liquid cloud water.
From the particle relabeling symmetry, a moist Kelvin circulation theorem is found, along with a moist PV conservation law that applies not for each individual parcel but for parcel-integrated PV, integrated over certain local volumes.





\keywords{moist atmospheric flows, clouds, Hamiltonian formulation, particle relabeling symmetry, Kelvin's circulation theorem, Ertel potential vorticity}
\end{abstract}


\section{Introduction}


Potential vorticity (PV) is one of the fundamental conserved quantities of geophysical fluid dynamics.
Conservation of PV was recognized long ago by \citet{rossby1939relation} and \citet{ertel1942neuer}. 
One important property of PV is that it is a material invariant, meaning that it
is conserved along fluid particle trajectories; this allows it to function as a tracer of fluid particles. This material invariance property is exploited in isentropic PV maps which have been used extensively as an operational diagnostic tool in the development of mid-latitude synoptic weather systems \citep[][]{hoskins1985use, hoskins1988potential, hoskins1991towards, thorpe1985diagnosis}.
Another important property is its invertibility: the instantaneous global distribution of PV with appropriate boundary conditions can be used to recover the balanced parts of the wind and temperature fields. 

  While the concept of PV is well-established for a dry atmosphere,
  the concept of PV is less clear for a moist atmosphere
  with clouds, phase changes, and latent heating.
  Several definitions of moist PV have been proposed, but a moist counterpart that retains all of the desirable properties of dry PV has been elusive. For example, one moist PV definition (here called $PV_v$) is defined in terms of the virtual potential temperature $\theta_v$, and it was shown to possess an invertibility principle which produces certain velocity and temperature fields \citep{schubert2001potential}. However, \citet{wetzel2020potential} have shown that $PV_v$ is not balanced and therefore does not recover the balanced velocity and temperature fields.
  Although $PV_v$ is not balanced, it is a conserved quantity in unsaturated regions and therefore can be used to monitor and diagnose latent heating in the atmosphere \citep[][]{davis1991potential, lackmann2002cold, gao2004generation, brennan2005influence, martin2013mid, brennan2008potential, lackmann2011midlatitude, madonna2014warm,bueler2017potential}. Another widely used PV (here called $PV_e$) is based on the equivalent potential temperature $\theta_e$ \citep[][]{bennetts1979conditional,emanuel1979inertial}.  One criticism of $PV_e$ is that it fails to possess an invertibility principle in the presence of phase changes \citep[][]{schubert2001potential, cao1995generation}. However, taking advantage of one or more additional balanced quantities $M$ involving water, extended $PV$--$M$ inversion can indeed be used to solve for the balanced moist flow using $PV_e$ \citep[][]{smith2017precipitating,wssm19,wetzel2020potential}. As a general principle, for a moist system, a single moist PV variable by itself is not sufficient information to find the balanced flow components, and additional moisture variables (M) need to be retained \citep[][]{smith2017precipitating}.

Furthermore, the moist PV quantities that have been proposed
are lacking the fundamental conservation property of dry PV:
conservation along fluid parcel trajectories.
Some moist PV quantities are conserved in one phase---either
inside a cloud or outside a cloud---but none are conserved
along fluid parcel trajectories in the presence of clouds
and phase changes. Hence, there remains an open question:
For a moist system with clouds and phase changes,
is there a moist PV quantity that is a material invariant?

While past definitions of moist PV have been proposed based on
physical and mathematical intuition, there is in fact a
systematic approach to identifying conservation principles.
In Hamiltonian mechanics and Lagrangian mechanics,
conservation principles are related to symmetries in the action,
via Noether's theorem
\citep[][]{emmy1918invariante,hill1951hamilton, olver2000applications}.
Therefore, as a systematic route to identifying a moist PV
conservation law, one can first write the moist atmospheric 
equations in a Hamiltonian or Lagrangian formulation, 
and then look for symmetries in the action
or the Lagrangian density. 


The goal of the present paper is to take the systematic approach
described above.
To first define a Hamiltonian or Lagrangian formulation,
we use the piecewise-quadratic energy that was recently
identified for the moist Boussinesq equations with phase changes
\citep{marsico2019energy}.
As the symmetry of interest for moist PV, we will seek a
moist analog of the symmetry associated with dry PV conservation:
particle relabeling symmetry
\citep[][]{bretherton1970note,ripa1981symmetries, salmon1988hamiltonian, shepherd1990symmetries, muller1995ertel}.



 
 
 The present paper is another contribution of the use of
 the Lagrangian or Hamiltonian formulation for geophysical fluid dynamics. For example, in other applications, by attaching additional constraints to the Lagrangian/Hamiltonian, the Lagrangian/Hamiltonian formulation provides a systematic way of deriving approximate dynamical equations that retain analogs of exact conservation laws
 \citep[][]{salmon1998lectures, holm1998euler, cotter2014variational}.
 Well-known examples of such approximate models are the quasi-geostrophic, shallow water, KdV and Green-Naghdi equations \citep[see][and references therein]{salmon1988hamiltonian}. 
 Additionally, the Hamiltonian formulation is useful in nonlinear stability theory 
 \citep[][]{salmon1988hamiltonian}. Other applications include data assimilation \citep[][]{cotter2013data,hastermann2021balanced}, statistical analysis \citep[][]{abramov2003hamiltonian,majda2019statistical, moore2020anomalous} and stochastic partial differential equations (SPDEs) for fluid dynamics from  a stochastic variational principle (SVP) \citep[][]{holm2015variational}. It has also been used in the development of numerical methods \citep[][]{pavlov2011structure}.

In Section \ref{sec:background}, we provide background about the governing moist Boussinesq equations and their conservation of energy principle.  Next in Section \ref{sec:lagrangian-formulation}, we show that the energy is differentiable, allowing for a formulation of the dynamics from the perspective of Lagrangian mechanics.
We further demonstrate that Hamilton's principle leads to  Euler-Lagrange equations that are equivalent to the original equations.  Section \ref{sec:pv} explains how to adapt particle-relabeling symmetry to the moist system with phase changes, and thereby derive moist analogues of Kelvin's circulation theorem and conservation of potential vorticity.  Conclusions are given in Section \ref{sec:conclusions}.

\section{Background}
\label{sec:background}

As background, we describe the model equations
in section~\ref{subsec:governing-equations}
and the conserved energy in section~\ref{subsec:energy}.
The energy will be used later to suggest the form
of a Hamiltonian and a Lagrangian.

\subsection{Governing Equations}
\label{subsec:governing-equations}

We start by presenting the model equations, which are moist
Boussinesq equations with phase changes. 
This type of model has been used for many purposes 
and with varying degree of idealization
\citep[e.g.,][]{k61,s76,b87i,cd93,sby06,ps10,marsico2019energy,zss21jfm,zss21jnls}.
While a Boussinesq framework is used here,
one would suspect that similar results could be obtained
for an anelastic or compressible system,
although such cases are beyond the scope of the
present paper. 
A Boussinesq framework is used here instead of an anelastic
or compressible framework because it allows for
some calculations to be in simpler form and because
it allows use of an energy function
(section~\ref{subsec:energy})
that has a simple piecewise quadratic form
\citep{marsico2019energy}.

When the two
thermodynamic variables are chosen to be $\theta_e$ and $q_t$,
the equations take the form
\begin{align}
    \frac{D\vec{u}}{Dt} &= -\nabla \phi + b \hat{k}, \label{eqn:u-evol} \\
    \frac{D\theta_e}{Dt} + \frac{d\Tilde{\theta_e}}{dz}w &= 0, 
    \label{eqn:thetae-evol} \\
    \frac{Dq_t}{Dt} + \frac{d\Tilde{q_t}}{dz}w &= 0,
    \label{eqn:qt-evol} \\
    \nabla \cdot \vec{u} &=0,
    \label{eqn:incomp}
\end{align}
where $\vec{x} = (x,y,z)$ is the position vector, $\hat{k}$ is the unit vector in the $z$-direction,
$\vec{u} = (u,v,w)$ is the velocity vector,
$D/Dt=\partial/\partial t+\vec{u}\cdot\nabla$ is the material derivative,
$\phi = p'/\rho_0$, $p'$ is the pressure anomaly, $\rho_0$ is a constant background density,
 $\theta_e$ is the equivalent potential temperature anomaly, and $q_t$ is the anomalous total water mixing ratio. Every thermodynamic variable considered here has been decomposed into a background function of height $z$ and an anomalous part. For example, $\theta_e^{tot} = \Tilde{\theta}_e(z)+ \theta_e(\vec{x},t)$ and
$q_t^{tot} = \tilde{q}_t(z) + q_t(\vec{x},t)$. 
The background vertical gradients, $d\tilde{\theta}_e/dz$ and
$d\tilde{q}_t/dz$, will be assumed to be constants,
in analogy with a common setup of the Boussinesq equations
in the dry case \citep[e.g.,][]{m03}.

The buoyancy $b$ is influenced by phase changes
of water, and $b$ can be expressed as a function of $\theta_e$, $q_t$ and $z$. To do this, one can start from a definition of 
\begin{align}
     b &= g \left( \frac{\theta}{\theta_0} + R_{vd}q_v -q_l  \right),
     \label{eqn:b-def-theta}
\end{align}
where $\theta_0 \approx 300$ K is the constant background potential temperature, $g \approx 9.8 \mbox{ m s}^{-2}$ is the acceleration due to gravity and $R_{vd} = (R_v/R_d) - 1 \approx 0.61$, where $R_d$ is the gas constant for dry air and $R_v$ is the gas constant for water vapor.
The three variables in this buoyancy expression are
potential temperature $\theta$,
water vapor mixing ratio $q_v$,
and liquid water mixing ratio $q_l$,
and they can be related to $q_t$ and $\theta_e$
as described below.

The set of variables $(\theta,q_v,q_l)$
can be related to the set of variables $(\theta_e,q_t)$
in the following way.
The total water $q_t$ is defined as 
the sum of water vapor $q_v$ and liquid water $q_l$: 
\begin{equation}
    q_t=q_v+q_l.
    \label{eqn:qt-def}
\end{equation}
The second variable, equivalent potential temperature $\theta_e$,
is defined here in linearized form as 
\begin{equation}
    \theta_e = \theta + \frac{L_v}{c_p}q_v,
    \label{eqn:thetae-def}
\end{equation}
where the latent heat is a constant with value $L_v \approx 2.5 \times 10^6 J kg^{-1}$ and specific heat is a constant with value $c_p \approx 10^3 J kg^{-1}K^{-1}$.
From these definitions in (\ref{eqn:qt-def}) and (\ref{eqn:thetae-def}), 
one can find the variables $(\theta_e,q_t)$
if given the variables $(\theta,q_v,q_l)$.


For the opposite direction, if given $q_t$, one can
partition it into its vapor ($q_v$) and liquid ($q_l$) components
by using the saturation mixing ratio $q_{vs}$, which acts as 
a threshold. The saturation mixing ratio will be assumed to be a function of only $z$ by making the assumption that $T^{tot}$ and $p^{tot}$ are close to the background states $\Tilde{p}(z)$ and $\Tilde{T}(z)$, which in turn depend only on the height, in which case
$q_{vs}^{tot}(T^{tot},p^{tot})\approx q_{vs}^{tot}(\tilde{T}(z),\tilde{p}(z))$
\citep[see, e.g., the appendix of][]{hmss13}.
Two cases are possible, depending on whether
$q_t$ is below the threshold ($q_t<q_{vs}$) 
or above the threshold ($q_t>q_{vs}$).
If $q_t$ is below the threshold ($q_t<q_{vs}$),
then no liquid water is present (i.e., $q_l=0$),
and the water is all in the form of vapor: $q_t=q_v$.
On the other hand, 
if $q_t$ is above the threshold ($q_t>q_{vs}$),
then the water vapor is at its saturation value
(i.e., $q_v=q_{vs}$), and the remaining water is liquid water:
$q_l=q_t-q_{vs}$.
To encompass both of these cases in a unified way,
the formulas for $q_v$ and $q_l$ can be written as
\begin{align}
    q_v &= \min(q_t,q_{vs}),
    \label{eqn:qv-def} 
    \\
    q_l &= \max(0,q_t-q_{vs}).
    \label{eqn:ql-def}
\end{align}
The two cases will be called the unsaturated phase
(if $q_v<q_{vs}$) and the saturated phase (if $q_v=q_{vs}$).

If $\theta_e$ and $q_t$ are both given, then
one can also see that
\begin{equation}
    \theta=\theta_e-\frac{L_v}{c_p}\min(q_t,q_{vs}).
    \label{eqn:theta-def}
\end{equation}
This expression follows from 
(\ref{eqn:thetae-def}) and (\ref{eqn:qv-def}).
Together, the three expressions
(\ref{eqn:qv-def}), (\ref{eqn:ql-def}), and (\ref{eqn:theta-def})
give a specification  
of the variables $(\theta,q_v,q_l)$ 
if given the variables $(\theta_e,q_t)$.

Note that many of the expressions above have been written in 
terms of anomalies, such as $\theta_e$, rather than the
total expressions, such as
$\theta_e^{tot} = \Tilde{\theta}_e(z)+ \theta_e(\vec{x},t)$,
which include background states.
In particular, notice that the fundamental condition for
saturation should be $q_t^{tot}=q_{vs}^{tot}$.
Nevertheless, one can choose the background state to be
unsaturated with $\tilde{q}_t(z)=\tilde{q}_v(z)$,
and one can also break $q_{vs}^{tot}$ into background
and anomaly as $q_{vs}^{tot}=\tilde{q}_{vs}+q_{vs}$,
with the choice of $\tilde{q}_{vs}(z)=\tilde{q}_{v}(z)$.
With these choices of
$\tilde{q}_t(z)=\tilde{q}_{v}(z)=\tilde{q}_{vs}(z)$,
it follows that the saturation condition
$q_t^{tot}=q_{vs}^{tot}$ can be written
in terms of anomalies in a simple way, as $q_t=q_{vs}$. 

It is convenient to rewrite these equations by replacing the thermodynamic variables, $\theta_e$ and $q_t$, with the unsaturated and saturated buoyancy variables, $b_u$ and $b_s$. The variables $b_u$ and $b_s$ are defined so that the buoyancy $b$ can be written simply as
\begin{align}
    b = b_u H_u + b_s H_s,
\end{align}
where $H_u$ and $H_s$ are Heaviside functions that are indicators
of the unsaturated and saturated regions, respectively:
\begin{equation}
    H_u = 
    \begin{cases}
    1 & \text{if}\ q_t < q_{vs}(z) \\
    0 & \text{if}\ q_t \ge q_{vs}(z),
    \end{cases}
    \qquad
    H_s = 1-H_u.
\label{def:indicatorqt}
\end{equation}
To define $b_u$ and $b_s$, the buoyancy definition in
(\ref{eqn:b-def-theta}) is rewritten in terms of equivalent potential
temperature $\theta_e$, total water $q_t$, and prescribed saturation
mixing ratio $q_{vs}$, using (\ref{eqn:qt-def})--(\ref{eqn:theta-def}),
which yields
\begin{align}
    b_u &= g \left( \frac{\theta_e}{\theta_0} + \left( R_{vd} - \frac{L_v}{c_p\theta_0}\right)q_t \right), 
    \label{eqn:bu-def} \\
    b_s &= g \left( \frac{\theta_e}{\theta_0} + \left( R_{vd} - \frac{L_v}{c_p\theta_0} + 1 \right)q_{vs}-q_t \right).
    \label{eqn:bs-def}
\end{align}
These types of unsaturated and saturated buoyancy variables
have also been used in other work on moist convection
\citep[e.g.,][]{k61,b87i,ps10,smith2017precipitating}
and are sometimes called the dry and moist buoyancy variables.
Notice that $b=b_u$ in unsaturated regions, and $b=b_s$ in
saturated regions, but the variables $b_u$ and $b_s$ are defined
and exist everywhere, since they are defined as functions of
$\theta_e, q_t,$ and $q_{vs}$.  

The governing equations in terms of $b_u$ and $b_s$ are
\begin{align}
    \frac{D \vec{u}}{Dt} &= -\nabla \phi + (b_u H_u + b_s H_s) \hat{k} \label{eqn:u-evol-bubs} \\
    \frac{D b_u}{Dt} + N_u^2 w &= 0
    \label{eqn:bu-evol} \\
    \frac{D b_s}{Dt} + N_s^2 w &= 0,
    \label{eqn:bs-evol} \\
    \nabla \cdot \vec{u} &=0,
    \label{eqn:incomp-bubs}
\end{align} 
where the unsaturated and saturated buoyancy frequencies,
$N_u^2$ and $N_s^2$, are constants given by
\begin{align}
    N_u^2 = g \frac{d}{dz}\left( \frac{\Tilde{\theta}_e}{\theta_0} + \left( R_{vd} - \frac{L_v}{c_p\theta_0}\right)\Tilde{q}_t \right),
    \label{eqn:Nu-def} \\
    N_s^2 = g \frac{d}{dz}\left( \frac{\Tilde{\theta}_e}{\theta_0} - \left( R_{vd} - \frac{L_v}{c_p\theta_0}\right)q_{vs}(z) - \Tilde{q}_t \right).
    \label{eqn:Ns-def}
\end{align}
The evolution equations for $b_u$ and $b_s$
in (\ref{eqn:bu-evol})--(\ref{eqn:bs-evol})
follow from the evolution equations for $\theta_e$ and $q_t$
in (\ref{eqn:thetae-evol})--(\ref{eqn:qt-evol}),
using the definitions of $b_u$ and $b_s$ in
(\ref{eqn:bu-def})--(\ref{eqn:bs-def});
and the definitions of $N_u^2$ and $N_s^2$ in
(\ref{eqn:Nu-def})--(\ref{eqn:Ns-def}) also follow from
this derivation.
Finally, the indicator functions (\ref{def:indicatorqt}) can be written as
\begin{equation}
    H_u = 
    \begin{cases}
    1 & \text{if}\ b_u > b_s \\
    0 & \text{if}\ b_u \le b_s,
    \end{cases}
    \qquad
    H_s = 1-H_u.
\label{def:indicatorbubs}
\end{equation}
With this reformulation of the equations in terms of buoyancy variables,
many aspects of the subsequent discussion will be simplified.


\subsection{Piecewise Quadratic Energy}
\label{subsec:energy}


In subsequent sections, to explore a possible particle-relabeling
symmetry of the Lagrangian, it will first be necessary to define
a suitable Lagrangian functional. To help motivate the form of a
Lagrangian functional for the model in
(\ref{eqn:u-evol})--(\ref{eqn:incomp}),
we now describe an energy function that was recently derived
\citep{marsico2019energy}.

The model in (\ref{eqn:u-evol})--(\ref{eqn:incomp}) or
(\ref{eqn:u-evol-bubs})--(\ref{eqn:incomp-bubs})
has an associated energy that is given by
    \begin{align} \label{eq:energy}
        E =  
        {\cal{K}} + {\cal{V}} 
    = \frac{1}{2} \vec{u} \cdot \vec{u} + \frac{1}{2} \frac{b_u^2}{N_u^2} H_u +  \frac{1}{2} \frac{b_s^2}{N_s^2} H_s + \frac{1}{2} \frac{N_u^2 N_s^2}{N_u^2-N_s^2} M^2 H_u 
    \end{align} 
and evolves according to
\begin{equation}
    \frac{\partial E}{\partial t} + \nabla\cdot [\vec{u}(E+\phi)] = 0,
\end{equation}
so that the domain-integrated energy is conserved
\citep{marsico2019energy}.
In the energy definition in (\ref{eq:energy}),
the first term is the kinetic energy
${\cal{K}}$,
and the latter three terms are the potential energy ${\cal{V}}$.
The $b_u^2$ and $b_s^2$ terms are the buoyant potential energy
terms in the unsaturated and saturated phases, respectively.
The new variable $M$ that appears in the potential energy
is actually related to $b_u$ and $b_s$ via
\begin{equation}
    M = \frac{b_u}{N_u^2} - \frac{b_s}{N_s^2},
    \label{def:M}
\end{equation}
and it is a material invariant---i.e.,
\begin{equation}
    \frac{DM}{Dt}=0,
\end{equation}
which can be seen from (\ref{eqn:bu-evol}), (\ref{eqn:bs-evol}),
and (\ref{def:M}).
This variable $M$ is associated with the additional eigenmode that is present in moist systems as opposed to dry systems \citep{hernandez2015stability, smith2017precipitating}.
The $M^2$ term in the potential energy in (\ref{eq:energy})
is a moist energy term that accounts for the latent heat released during change of phase. Notice that each potential energy term
is multiplied by either $H_u$ or $H_s$, so the form of the
potential energy will be different in different phases.
In this way, the energy in (\ref{eq:energy}) is piecewise quadratic,
as opposed to the quadratic energy that arises in the dry case.


Since the energy in (\ref{eq:energy}) is piecewise-quadratic,
it is positive if the coefficients $N_u^2$, $N_s^2$, and
$N_u^2-N_s^2$ are positive.
Each of $N_u^2$ and $N_s^2$ will be positive if the fluid is
stably stratified in the dry and moist sense,
according to (\ref{eqn:Nu-def})--(\ref{eqn:Ns-def}).
For simpler expressions of $N_u^2$ and $N_s^2$,
approximate forms of (\ref{eqn:Nu-def})--(\ref{eqn:Ns-def}) are
\begin{equation}
    N_u^2 \approx \frac{g}{\theta_0}\frac{d\tilde{\theta}}{dz}, \qquad
    N_s^2 \approx \frac{g}{\theta_0}\frac{d\tilde{\theta}_e}{dz},
    \label{eqn:Nu-Ns-approx}
\end{equation}
which indicate that $N_u^2$ and $N_s^2$ will be positive
if the background vertical gradients of $\theta$ and $\theta_e$
are positive. In this case, the difference $N_u^2-N_s^2$ will be 
positive if $d\tilde{q}_v/dz$ is negative, as can be seen from
(\ref{eqn:thetae-def}) and (\ref{eqn:Nu-Ns-approx}).

While the energy in (\ref{eq:energy}) is piecewise-defined
in terms of Heaviside functions, it was shown by
\citet{marsico2019energy} that the energy is
continuous across phase boundaries. 
In other words, the potential energy in the unsaturated phase,
\begin{equation}
    \frac{1}{2} \frac{b_u^2}{N_u^2}  + \frac{1}{2} \frac{N_u^2 N_s^2}{N_u^2-N_s^2} M^2,
    \label{eqn:PEu}
\end{equation}
and the potential energy in the saturated phase,
\begin{equation}
    \frac{1}{2} \frac{b_s^2}{N_s^2},
    \label{eqn:PEs}
\end{equation}
will be equal at the phase interface. To see this, recall from
(\ref{def:indicatorbubs}) that the phase interface can be defined
as locations where $b_u=b_s$. Consequently, at the phase interface,
one can rewrite the potential energy from the unsaturated phase as
\begin{align}
    \frac{1}{2} \frac{b_u^2}{N_u^2}  + \frac{1}{2} \frac{N_u^2 N_s^2}{N_u^2-N_s^2} M^2
    &= 
    \frac{1}{2} \frac{b_u^2}{N_u^2}  + \frac{1}{2} \frac{N_u^2 N_s^2}{N_u^2-N_s^2} \left( \frac{b_u}{N_u^2} - \frac{b_s}{N_s^2} \right)^2
    \\
    &=
    \frac{1}{2} \frac{b_u^2}{N_u^2}  + \frac{1}{2} \frac{N_u^2 N_s^2}{N_u^2-N_s^2} \left( \frac{b_u^2}{N_u^4} - \frac{2b_u b_s}{N_u^2 N_s^2} + \frac{b_s^2}{N_s^4} \right)
    \\
    &=
    \frac{1}{2} \frac{b_s^2}{N_u^2}  + \frac{1}{2} \frac{N_u^2 N_s^2}{N_u^2-N_s^2} \left( \frac{b_s^2}{N_u^4} - \frac{2 b_s^2}{N_u^2 N_s^2} + \frac{b_s^2}{N_s^4} \right)
    \\
    &=
    \frac{1}{2} \frac{b_s^2}{N_s^2},
\end{align}
so that (\ref{eqn:PEu}) and (\ref{eqn:PEs}) are equal
at the phase interface,
where the definition of $M$ from (\ref{def:M}) was also used
in the calculation.
Intuitively, the buoyant energy $b_u^2/N_u^2$ should be smaller
than $b_s^2/N_s^2$ at the phase interface, since $N_u^2$ is
typically larger than $N_s^2$; the moist energy $M^2$ then
compensates for the difference and ensures continuity of 
potential energy.

Beyond this earlier demonstration of continuity, 
it will be shown in the present paper that the energy is
also continuously differentiable, which allows us to take
derivatives, which are needed for using the energy in a Hamiltonian or 
Lagrangian formulation.

\section{Lagrangian Formulation and Euler--Lagrange Equations}
\label{sec:lagrangian-formulation}

The goal of this section is to present a Lagrangian formulation
of the model in (\ref{eqn:u-evol})--(\ref{eqn:incomp}) or
(\ref{eqn:u-evol-bubs})--(\ref{eqn:incomp-bubs}).
Then, in section~\ref{sec:pv}, the Lagrangian formulation
can be used to investigate potential vorticity and
particle-relabeling symmetry of the Lagrangian.

\subsection{Lagrangian density and its regularity at cloud edge}
\label{sec:densityregularity}

The piecewise quadratic energy in (\ref{eq:energy}) can be used to construct a Lagrangian density $\ell$ for our system as the difference between kinetic ${\cal{K}}$ and potential ${\cal{V}}$ energy terms:
     \begin{align} 
\label{eq:lagrangian_density}
\ell = 
{\cal{K}} - {\cal{V}} 
= \frac{1}{2} \vec{u} \cdot \vec{u} - \biggl (\frac{1}{2} \frac{b_u^2}{N_u^2} H_u +  \frac{1}{2} \frac{b_s^2}{N_s^2} H_s + \frac{1}{2} \frac{N_u^2 N_s^2}{N_u^2-N_s^2} M^2 H_u 
\biggr),
\end{align}
where the potential energy $\cal{V}$ was discussed above
in section~\ref{subsec:energy}.
For instance, while $\cal{V}$ is defined in a piecewise manner
in unsaturated and saturated phases, it is nevertheless a
continuous function.


In order to proceed with a Lagrangian formulation of the dynamics including phase changes, it is necessary to demonstrate that the density $\ell$ given by (\ref{eq:lagrangian_density}) is differentiable at phase boundaries, which are indicated by the Heaviside functions $H_u, H_s$.  Thus we proceed to compute derivatives in the unsaturated ($b_u > b_s$) and saturated ($b_u < b_s$) phases separately, and then to show equality of their one-sided limits as the phase boundary ($b_u=b_s$) is approached.

Let us fist consider (\ref{eq:lagrangian_density}) in an unsaturated flow region with $H_u=1$ and $H_s=0$, and 
denote $q$ as a placeholder for $x,y,z$ or $t$.  A derivative with respect to $q$ of (\ref{eq:lagrangian_density}) in the unsaturated phase gives
\begin{align}
\frac{\partial \ell}{\partial q} &= \vec{u} \cdot \frac{\partial \vec{u}}{\partial q} - \frac{b_u}{N_u^2} \frac{\partial b_u}{\partial q} - \frac{N_u^2 N_s^2}{N_u^2-N_s^2} M \frac{\partial M}{\partial q}\\
&=\vec{u} \cdot \frac{\partial \vec{u}}{\partial q}- \frac{b_u}{N_u^2} \frac{\partial b_u}{\partial q} - \frac{1}{N_u^2-N_s^2} \left( b_u \frac{N_s^2}{N_u^2}\frac{\partial b_u}{\partial q} - b_u\frac{\partial b_s}{\partial q} + b_s \frac{N_u^2}{N_s^2}\frac{\partial b_s}{\partial q} - b_s\frac{\partial b_u}{\partial q}\right), 
\label{eqn:dldqunsat}
\end{align}
where we have used the definition of $M$ given by (\ref{def:M}).
As the phase interface is approached from the unsaturated side, $b_u$ tends to $b_s$ from above, and therefore
\begin{align}
\lim_{b_u-b_s \to 0^+} \frac{\partial \ell}{\partial q} &= 
\lim_{b_u-b_s \to 0^+} \left (
\vec{u} \cdot \frac{\partial \vec{u}}{\partial q}- \left(\frac{ [(N_u^2-N_s^2) + N_s^2]b_u - N_u^2 b_s}{N_u^2(N_u^2-N_s^2)} \right) \frac{\partial b_u}{\partial q} - \left ( \frac{b_s N_u^2 - b_u N_s^2}{N_s^2(N_u^2-N_s^2)}   \right)\frac{\partial b_s}{\partial q} \right )\\
&=\vec{u} \cdot \frac{\partial \vec{u}}{\partial q} -  \frac{b_s}{N_s^2} \frac{\partial b_s}{\partial q}. 
\label{eqn:limdldqunsat}
\end{align}
Starting again from (\ref{eq:lagrangian_density}), but now in a saturated region, the derivative of $\ell$ with respect to $q$ is
\begin{align}
\frac{\partial \ell}{\partial q} &= \vec{u} \cdot \frac{\partial \vec{u}}{\partial q} - \frac{b_s}{N_s^2} \frac{\partial b_s}{\partial q}.
\label{eqn:dldqsat}
\end{align}
From (\ref{eqn:limdldqunsat}) and (\ref{eqn:dldqsat}), one can see that \begin{equation}
\lim_{b_u-b_s \to 0^+} \frac{\partial \ell}{\partial q} = 
\lim_{b_s-b_u  \to 0^-} \frac{\partial \ell}{\partial q} = \vec{u} \cdot \frac{\partial \vec{u}}{\partial q} -  \frac{b_s}{N_s^2} \frac{\partial b_s}{\partial q},
\label{eqn:limdldqequal}
\end{equation}
which establishes differentiability of the Lagrangian density $\ell$ at phase boundaries where $b_u=b_s$.

More generally, while the derivatives in 
(\ref{eqn:dldqunsat}) and (\ref{eqn:dldqsat}) were calculated 
by restricting attention to unsaturated and saturated regions, respectively, one might further desire a derivative formula
that is valid throughout the entire domain.
In this direction,
and for a generic scenario, consider a piecewise-defined
function $g(q)$ given by
\begin{equation}
    g(q) = g_u(q) H_u + g_s(q) H_s.
    \label{eqn:general-g}
\end{equation}
For example, the Lagrangian density in (\ref{eq:lagrangian_density})
is of this form.
Then its derivative can be calculated as
\begin{align}
    \frac{dg}{dq} 
    &= 
    \frac{d}{dq}\left[ g_u(q) H_u + g_s(q) H_s \right]
    \\
    &= g_u'(q) H_u + g_s'(q) H_s
    +g_u(q) \frac{dH_u}{dq} + g_s(q) \frac{dH_s}{dq} 
    \\
    &= g_u'(q) H_u + g_s'(q) H_s
    +g_u(q) \frac{dH_u}{dq} + g_s(q) \frac{d}{dq} (1-H_u)
    \\
    &= g_u'(q) H_u + g_s'(q) H_s
    +[g_u(q)-g_s(q)] \frac{dH_u}{dq} 
    \\
    &= g_u'(q) H_u + g_s'(q) H_s,
\end{align}
where the last equality follows from an additional assumption that
$g$ is continuous,
which implies that $g_u(q)-g_s(q)=0$ at the phase interface,
so that the coefficient of the Dirac-delta $dH_u/dq$ above
is zero at the phase interface.
Consequently, in calculating the derivative of a piecewise-defined
function as in (\ref{eqn:general-g}), if the function is also continuous,
then one can simply use
\begin{equation}
\label{eqn:dgdq}
    \frac{dg}{dq} = g_u'(q) H_u + g_s'(q) H_s,
\end{equation}
by differentiating the coefficients of $H_u$ and $H_s$,
without needing to consider derivatives of $H_u$ and $H_s$ 
themselves.  We will use this result below in section \ref{sec:hamiltonsprinciple}.

\subsection{Hamilton's Principle and the Euler--Lagrange Equations}
\label{sec:hamiltonsprinciple}

The $\ell$ defined in (\ref{eq:lagrangian_density})
is a candidate for a Lagrangian density,
which can potentially be used to define an associated set of
Euler--Lagrange equations, via a variational principle---i.e.,
the principle of least action, or Hamilton's principle.
A priori, it is not immediately clear whether this will work here,
due to complications of phase changes.
In section~\ref{sec:densityregularity} above, 
we established that, at the very least,
$\ell$ is differentiable, so it is at least possible to take
variational derivatives of $\ell$.
Now we investigate whether a variational principle, using
$\ell$ as the Lagrangian density, will lead to the model in 
(\ref{eqn:u-evol})--(\ref{eqn:incomp}) or
(\ref{eqn:u-evol-bubs})--(\ref{eqn:incomp-bubs})
as the associated Euler--Lagrange equations.

Hamilton's principle states that the dynamics of a system are determined according to a variational statement, and for fluid systems, the variational statement may be expressed in either the 
Lagrangian\footnote{Note that the adjective \emph{Lagrangian} is sometimes used in the sense of Lagrangian versus Eulerian description of fluid dynamics, and sometimes in the sense of Lagrangian versus Hamiltonian formulation of classical mechanics. The different uses here should be clear from context.}
description or the Eulerian description \citep{hill1951hamilton,bretherton1970note,salmon1982hamilton,salmon1988hamiltonian,salmon1998lectures,cotter2014variational}. Our goal is to derive the model (\ref{eqn:u-evol-bubs})--(\ref{eqn:incomp-bubs}), which is written in the Eulerian description. Nevertheless, for clarity and for further use below, we shall start now from the Lagrangian description and then briefly recall the steps in the transformation from Lagrangian to Eulerian description.
In the Lagrangian description, one considers variations $\delta\vec{x}$ in the particle trajectories $\vec{x} = \vec{x}(\vec{a},t)$ corresponding to particles labeled by $\vec{a}$.
In the Eulerian description, the system is described by the velocity $\vec{u}$ at fixed locations, and thus one must use the inverse flow map $\vec{a} = \vec{a}(\vec{x},t)$ (assumed one-to-one) to transform the variational statement accordingly.
As in the previous sections, it will be convenient to work in Cartesian coordinates for simplicity.

Following fluid particles labeled by $\vec{a}$, the Lagrangian $\mathcal{L}$ for the system is given by
\begin{align}
\mathcal{L}(t)
    = 
    \iiint d V_a \;
    \ell(\vec{x}(\vec{a},t),\vec{x}_t(\vec{a},t),\vec{a})= \iiint 
    d V_a \;
    \frac{1}{2}\left (  
    \vec{x}_t \cdot \vec{x}_t - 
    \frac{b_u^2}{N_u^2} H_u - \frac{b_s^2}{N_s^2} H_s -  \frac{N_u^2 N_s^2}{N_u^2-N_s^2} M^2 H_u \right) 
    \label{eq:lagrangian_langrangian}
\end{align}
where $\vec{a}$ is chosen such that $dV_a$ represents a differential mass element, $dm$,  in particle-label space. For example, $\vec{a}$ can be chosen as the initial position $\vec{x}_0$. We make use of a different such choice of $\vec{a}$ in later sections. The notation $\vec{x}_t$ means the derivative of $\vec{x}$ with respect to time $t,$ keeping particle label $\vec{a}$ fixed.  
Hamilton's principle is stated in terms of the action 
$\mathcal{A}$:
\begin{equation}
    \delta \mathcal{A} = 0,\quad \mathcal{A} = \int_{t_1}^{t_2} dt \mathcal{L}(t), 
    \label{eq:hamiltonsprinciplelagrange}
\end{equation}
for variations $\delta \vec{x}(\vec{a},t)$ that vanish outside the range $t\in[t_1,t_2]$.  These variations must also tend to zero as 
$\vec{x}$ tends to infinity, or satisfy no-flow through the domain boundary
\citep[see for example][]{bretherton1970note}.
To transform (\ref{eq:hamiltonsprinciplelagrange}) to the Eulerian description, one must change variables in the volume element using
\begin{equation}
dV_a =   {\cal{J}}dV_x = \frac{\partial (\vec{a})}{\partial (\vec{x})} dV_x,
    \label{def:determinantjacobian}
    \end{equation}
where $dV_x$ is the volume element in physical space, ${\cal J}$ is the Jacobian determinant, and we adopt the notation ${\cal  J} \equiv \partial(\vec{a})/\partial (\vec{x})$ as in some of the previous literature.
In Cartesian coordinates, ${\cal J}$ is the determinant of the matrix $\partial a_i/\partial x_j$.
For labeling coordinates chosen such that equal volumes in $\vec{a}$-space have equal masses, 
$\partial(\vec{a})/\partial (\vec{x})$ is the mass density $\rho$.  Here we wish to describe incompressible Boussinesq dynamics, and thus we take the constant $\rho=1$.  Using the Eulerian notation $\vec{u} = \vec{x}_t$, Hamilton's principle (\ref{eq:hamiltonsprinciplelagrange}) becomes
\begin{equation}
    \delta \mathcal{A} = 0,\quad \mathcal{A} = \int_{t_1}^{t_2} dt  \iiint dV_x \; 
    \frac{1}{2} \left ( \vec{u} \cdot \vec{u} -  \frac{b_u^2}{N_u^2} H_u -   \frac{b_s^2}{N_s^2} H_s -  \frac{N_u^2 N_s^2}{N_u^2-N_s^2} M^2 H_u  \right),
    \label{eq:hamiltonsprincipleeulerian}
\end{equation}
where now the action $\cal{A}$ must be stationary with respect to arbitrary variations $\delta \vec{a}$.
Thus one must express the integrand as
$\ell = \ell(\vec{a}(\vec{x},t),...,\vec{x})$, where the $...$ means derivatives of $\vec{a}$ with respect to $\vec{x}$ and/or $t$.

For example, let us consider how to express the velocity $\vec{u}$ in (\ref{eq:hamiltonsprincipleeulerian}) as first derivatives of $\vec{a}$ with respect to $\vec{x}$ and $t$.  The latter can be achieved from conservation of label $\vec{a}$ following fluid particles:
\begin{equation}
    \frac{D}{Dt}\vec{a}  = \vec{a}_t + \vec{u} \cdot \nabla \vec{a}= 0,
\label{eq:aconservation}
\end{equation}
by solving (\ref{eq:aconservation}) as a $3\times 3$ system for $\vec{u}$ \citep{salmon1988hamiltonian}, and we note that the $\nabla$ without a subscript is the gradient operator $\nabla = (\partial_x,\partial_y,\partial_z)$.  Alternatively, an equivalent variational statement is found by appending
(\ref{eq:aconservation}) to 
(\ref{eq:hamiltonsprincipleeulerian}) as constraints (the Lin constraints), and then 
considering independent variations $\delta\vec{u}$ and $\delta\vec{a}$
\citep{bretherton1970note,salmon1988hamiltonian}.
In addition to the Lin constraints, a constraint is required to enforce the incompressibility condition $\nabla \cdot \vec{u}= 0$, thus leading to 
\begin{align}
    \mathcal{A} = \int dt \iiint dV_x  \; \biggl ( \frac{1}{2} \vec{u} \cdot \vec{u} - \frac{1}{2} \frac{b_u^2}{N_u^2} H_u -  \frac{1}{2} \frac{b_s^2}{N_s^2} H_s - \frac{1}{2} \frac{N_u^2 N_s^2}{N_u^2-N_s^2} M^2 H_u - \vec{\alpha}\cdot\frac{D \vec{a} }{Dt} -  \lambda (\nabla \cdot u) \biggr),
\end{align}
where $\lambda$, $\vec{\alpha} = (\alpha,\beta,\gamma)$ are Lagrange multipliers.
As will be shown below, the multiplier $\lambda$ contributes to the pressure. 
Next, the potential energy in the integrand can be re-written in terms of the material invariants $b_u^{tot} = b_u^{tot}(\vec{a})$ and $b_s^{tot} = b_s^{tot}(\vec{a})$ using the relations
\begin{align}\label{eq:butotbstot}
        b_u &= b_u^{tot} - N_u^2z,\quad 
        b_s = b_s^{tot} - N_s^2z.
        \end{align}
One can verify that 
$b_u^{tot},b_s^{tot}$ are invariant following fluid particles according to the conservation equations
\begin{equation}
    \frac{D}{Dt}b_u^{tot} = 
    \frac{D}{Dt}b_s^{tot} = 0,
    \quad \frac{D}{Dt} = \frac{\partial}{\partial t} + \vec{u} \cdot \nabla.
    \label{eq:consbutotbstot}
\end{equation}
Using (\ref{eq:butotbstot}), one can also re-write 
$H_u, H_s$, and $M$ in the potential energy to be functions of $b_u^{tot}, b_s^{tot}$,
based on their definitions in (\ref{def:indicatorbubs}) and (\ref{def:M}).
Notice that, via (\ref{eq:butotbstot}), the height $z$ now appears explicitly in the potential energy terms, along with $b_u^{tot},b_s^{tot}$.
However, the explicit $z$-dependence does not have an influence on
variations $\delta\vec{a}$ for the Lagrangian density 
$\ell = \ell(\vec{a}(\vec{x},t),...,\vec{x})$,
since the explicit $z$-dependence is part of 
the explicit $\vec{x}$-dependence
in $\ell = \ell(\vec{a}(\vec{x},t),...,\vec{x})$
and is separate from the \textit{implicit} $\vec{x}$-dependence
of the $\vec{a}(\vec{x},t)$-dependence of $\ell$.
As some physical connection, note that the potential energy
is now considered as a function of the three variables
$b_u^{tot},b_s^{tot},z$,
and the presence of these three variables is
analogous to the presence of three thermodynamic
variables $\theta_e^{tot},q_t^{tot},z$ or 
$s^{tot},q_t^{tot},p$
that are needed to describe the thermodynamic state of a moist system.
The quantities $b_u^{tot}$ and $b_s^{tot}$ are material invariants that
play the same type of role as the material invariants of
entropy $s^{tot}$ and total water mixing ratio $q_t^{tot}$,
and $z$ plays the role of pressure $p$ in
Boussinesq and anelastic systems \citep{p08}.
We continue to write the potential energy in terms of $b_u,b_s,M$ for simplicity of the expression; however, the transformation to the variables $b_u^{tot},b_s^{tot},z$ is henceforth implied.
Finally, we can exploit the material invariance of the buoyancy variables $b_u^{tot}$, $b_s^{tot}$ by assigning them as two of the particle labels:
$\vec{a} = (a,b,c)=(a,b_u^{tot},b_s^{tot})$,
thereby allowing for independent variations $\delta b_u^{tot}$ and $\delta b_s^{tot}$. Using the new labels, and integration by parts in the 
incompressibility constraint, we arrive at
\begin{align}
    \mathcal{A} = \int dt \iiint dV_x \left ( \frac{1}{2} \vec{u} \cdot \vec{u}  - \frac{1}{2} \frac{b_u^2}{N_u^2} H_u -  \frac{1}{2} \frac{b_s^2}{N_s^2} H_s - \frac{1}{2} \frac{N_u^2 N_s^2}{N_u^2-N_s^2} M^2 H_u - \alpha  \frac{Da}{Dt} - \beta \frac{Db_u^{tot}}{Dt} - \gamma \frac{Db_s^{tot}}{Dt} - \vec{u} \cdot \nabla \lambda \right ).
\end{align}
After all of the preceding transformations,
Hamilton's principle requires that the $\mathcal{A}$ be stationary with respect to variations in  $\vec{u}(\vec{x},t)$, $a(\vec{x},t)$, $b_u^{tot}(\vec{x},t)$, $b_s^{tot}(\vec{x},t)$, $\alpha (\vec{x},t)$, $\beta (\vec{x},t)$, $\gamma (\vec{x},t)$ and $\lambda (\vec{x},t)$. These variations result in the following:
\begin{align}
    &\delta u : \vec{u} - \alpha\nabla a - \beta \nabla b_u^{tot} - \gamma \nabla b_s^{tot} - \nabla \lambda = 0,
    \label{eqn:variationu}\\
    &\delta \alpha : \frac{Da}{Dt} = 0, 
    \label{eqn:variationA}\\
    &\delta a: \frac{D\alpha}{Dt} = 0, \\
    &\delta \beta : \frac{Db_u^{tot}}{Dt} = 0, \label{eqn:variationB}\\
    &\delta b_u^{tot}: \frac{D\beta}{Dt} - \frac{b_u}{N_u^2}  H_u -    \frac{N_s^2}{N_u^2-N_s^2} M H_u =0, \\
    &\delta \gamma : \frac{Db_s^{tot}}{Dt} = 0, \label{eqn:variationC}\\
    &\delta b_s^{tot}: \frac{D\gamma}{Dt} - \frac{b_s}{N_s^2} H_s + \frac{N_u^2}{N_u^2-N_s^2} M H_u =0, \\
    &\delta \lambda : \nabla \cdot \vec{u} = 0,
    \label{eqn:variationphi}
\end{align}
where we have used (\ref{eqn:dgdq}).
To obtain the momentum equation, one can compute $D \vec{u}/D t$ using (\ref{eqn:variationu}), and then substitute from the remaining equations in (\ref{eqn:variationA})-(\ref{eqn:variationphi}) to obtain
\begin{align}
    \frac{D\vec{u}}{Dt} &= \frac{D}{Dt} \left ( \alpha \nabla a + \beta \nabla b_u^{tot} + \gamma \nabla b_s^{tot} + \nabla \lambda \right) 
     = \frac{\partial}{\partial t} \left ( \alpha \nabla a + \beta \nabla b_u^{tot} + \gamma \nabla b_s^{tot} + \nabla \lambda \right) + \vec{u} \cdot \nabla \vec{u} \label{eqn:u1} \\
    &= \alpha \frac{\partial \nabla a}{\partial t} + \beta \frac{\partial \nabla b_u^{tot}}{\partial t} + \gamma \frac{\partial \nabla b_s^{tot}}{\partial t} + \frac{\partial \nabla \lambda}{\partial t} + \nabla a \frac{\partial \alpha}{\partial t} + \nabla b_u^{tot} \frac{\partial \beta}{\partial t} + \nabla b_s^{tot} \frac{\partial \gamma }{\partial t} + \vec{u} \cdot \nabla \vec{u} \label{eqn:u2} \\
    &= - \left( \alpha \nabla (\vec{u} \cdot \nabla a) + \beta \nabla (\vec{u} \cdot \nabla b_u^{tot}) + \gamma \nabla (\vec{u} \cdot \nabla b_s^{tot}) + \nabla (\vec{u} \cdot \nabla \lambda) \right) \nonumber \\
    &+ \nabla ( \vec{u} \cdot \nabla \lambda + \frac{\partial \lambda}{\partial t}) + H_u \frac{b_u}{N_u^2} \nabla b_u^{tot} + H_u \frac{N_s^2}{N_u^2-N_s^2} M \nabla b_u^{tot} + H_s \frac{b_s}{N_s^2} \nabla b_s^{tot} - H_u \frac{N_u^2}{N_u^2-N_s^2} M \nabla b_s^{tot}  \nonumber \\ 
    &+\vec{u} \cdot \nabla \vec{u} 
    - \left( \nabla a (\vec{u} \cdot \nabla \alpha) + \nabla b_u^{tot} (\vec{u} \cdot \nabla \beta) + \nabla b_s^{tot} (\vec{u} \cdot \nabla \gamma) \right) \label{eqn:u3} \\
    &= - \left( \alpha \nabla (\vec{u} \cdot \nabla a) + \beta \nabla (\vec{u} \cdot \nabla b_u^{tot}) + \gamma \nabla (\vec{u} \cdot \nabla b_s^{tot}) + \nabla (\vec{u} \cdot \nabla \lambda) \right) \nonumber \\
    &- \nabla \left( -\vec{u} \cdot \nabla \lambda - \frac{\partial \lambda}{\partial t} - \frac{1}{2}\frac{(b_u)^2}{N_u^2}H_u - \frac{1}{2}\frac{(b_s)^2}{N_s^2}H_s - \frac{1}{2} \frac{N_u^2 N_s^2}{N_u^2-N_s^2} M^2 H_u \right) + (b_u H_u + b_s H_s)\hat{k} + \vec{u} \cdot \nabla \vec{u} \nonumber \\
    &- \left( \nabla a (\vec{u} \cdot \nabla \alpha) + \nabla b_u^{tot} (\vec{u} \cdot \nabla \beta) + \nabla b_s^{tot} (\vec{u} \cdot \nabla \gamma) \right). \label{eqn:u4}
\end{align}
In moving from (\ref{eqn:u2}) to (\ref{eqn:u3}),
the term $\nabla (\vec{u} \cdot \nabla \lambda)$ was added and subtracted,
and several of the relations from
(\ref{eqn:variationA})-(\ref{eqn:variationphi})
were used to replace $\partial/\partial t$ terms.
Then, in moving from (\ref{eqn:u3}) to (\ref{eqn:u4}),
the term $\nabla b_u^{tot}$ was split into contributions from
$b_u$ and $N_u^2 z$, using the definition in (\ref{eq:butotbstot}),
and similarly for $\nabla b_s^{tot}$, leading to the
buoyancy term $(b_u H_u + b_s H_s)\hat{k}$ in (\ref{eqn:u4}).
Also used in moving from (\ref{eqn:u3}) to (\ref{eqn:u4})
are (\ref{def:M}), (\ref{eqn:limdldqequal}), and (\ref{eqn:dgdq}).
To proceed further, the main challenge is to identify and collect gradient terms so as to determine an effective pressure to recover the requisite governing equations.  With that goal, we use the identity
\begin{equation}
    \alpha \nabla (\vec{u} \cdot \nabla a) + \nabla a (\vec{u} \cdot \nabla \alpha) = \nabla (\vec{u} \cdot (\alpha\nabla a)) + \vec{u} \cdot \nabla (\alpha \nabla a) - [\nabla (\alpha\nabla a)]^T \vec{u} \\
\end{equation}
where $[\nabla (\alpha\nabla a)]^T \vec{u}$ is a matrix--vector 
multiplication that can be written in index notation as
$u_j\partial_{x_i}(\alpha \partial_{x_j} a)$,
with summation over repeated indices.
Using this identity in (\ref{eqn:u4}), we obtain
\begin{align}
\label{eqn:u-evol-bubs-variational}
    \frac{D\vec{u}}{Dt} &= -\nabla p + (b_u H_u + b_s H_s)\hat{k},
\end{align}    
where the pressure $p$ is related to the Lagrange multiplier $\lambda$ via
\begin{align} \label{eq:pressure}
    p &= \left( \frac{1}{2}|\vec{u}|^2 -\vec{u} \cdot \nabla \lambda - \frac{\partial \lambda}{\partial t} - \frac{1}{2}\frac{(b_u)^2}{N_u^2}H_u - \frac{1}{2}\frac{(b_s)^2}{N_s^2}H_s - \frac{1}{2} \frac{N_u^2 N_s^2}{N_u^2-N_s^2} M^2 H_u \right)\\
    &= \biggl (
    \ell -\vec{u} \cdot \nabla \lambda - \frac{\partial \lambda}{\partial t} \biggr ).
\end{align} 
Therefore we have demonstrated that the governing equations 
(\ref{eqn:variationB}),
(\ref{eqn:variationC}),
(\ref{eqn:variationphi}),
(\ref{eqn:u-evol-bubs-variational})
are obtained using Hamilton's principle, starting from the Lagrangian density (\ref{eq:lagrangian_density}) [compare to 
(\ref{eqn:u-evol-bubs})-(\ref{eqn:incomp-bubs}) and use (\ref{eq:butotbstot})].

\section{Circulation and Potential Vorticity in Moist Systems}
\label{sec:pv}

Using the Lagrangian formulation from
section~\ref{sec:lagrangian-formulation},
we can now investigate potential vorticity from the perspective
of particle-relabeling symmetry of the Lagrangian.
In this section, a moist version of
particle-relabeling symmetry is studied, and it is used to
identify various versions of potential vorticity conservation.

\subsection{Background on Noether's Theorem and Possible Outcomes}

As a systematic approach to seeking conservation laws, 
one could identify symmetries of the Lagrangian function
and apply Noether's theorem to arrive at a corresponding
conservation law \citep{emmy1918invariante}. 
For conservation of potential vorticity
and Kelvin's circulation theorem, the symmetry of interest
for dry dynamics is the
particle-relabeling symmetry
\citep[e.g.,][]{padhye1996relabeling,bretherton1970note,salmon1988hamiltonian}.
Here, we now seek a particle-relabeling
symmetry that is valid for moist dynamics with phase changes.
By taking such a systematic approach, we can discover statements for conservation of circulation and moist
potential vorticity in a system with phase changes.

Before considering particle-relabeling symmetry and
the systematic approach itself, it is worthwhile to
consider what the possible outcomes may be.
As examples, consider other known symmetries and their
associated conservation laws. One well-known symmetry is
translation-invariance, whereby the action is invariant
under translations of the $\vec{x}$ coordinates.
The corresponding conservation law is conservation of momentum.
A second symmetry is the particle-relabeling symmetry, which,
for a dry atmosphere, is known to correspond with
conservation of PV.
As examples of conservation laws, these two examples differ
in that it is the total (or global) momentum, integrated over
all fluid parcels, that is conserved, whereas the PV
is conserved for each individual fluid parcel.
A priori, for the moist case of the present paper, either scenario is
possible. In other words, it is possible that
moist PV will be conserved for each individual fluid parcel,
or that moist PV is globally conserved 
(upon integrating over all fluid parcels),
or some other type of conservation law.
By finding the moist particle-relabeling symmetry
and applying Noether's theorem, we will discover the
corresponding conservation law that arises.

\subsection{Particle-Relabeling Symmetry and PV}

In this section, our goal is to find a particle-relabeling symmetry for the action $\cal{A}$ defined in (\ref{eq:lagrangian_langrangian})-(\ref{eq:hamiltonsprinciplelagrange}), whose Euler-Lagrange equations are the moist system (\ref{eqn:u-evol-bubs})-(\ref{eqn:incomp-bubs}) with phase changes. 
The action for our system may be written as
 \begin{align}
\mathcal{A}
    = \int dt \iiint d V_a \; \ell(\vec{x}(\vec{a},t),\vec{x}_t(\vec{a},t),\vec{a})= \int dt \iiint d V_a \; 
    \frac{1}{2} \vec{x}_t \cdot \vec{x}_t - {\cal{V}}(b_u^{tot}(\vec{a}),b_s^{tot}(\vec{a}),z)
    \label{eq:action}
\end{align}
where $\vec{x} = \vec{x}(\vec{a},t)$ is the particle path, $\vec{x}_t = \vec{x}_t(\vec{a},t)$ is the fluid velocity following fluid particles,
and $b_u^{tot},b_s^{tot}$ are the total buoyancies specified by (\ref{eq:butotbstot})-(\ref{eq:consbutotbstot}).  Henceforth it is understood that the limits of integration and boundary conditions are appropriately defined, as mentioned in section \ref{sec:hamiltonsprinciple} and discussed by \cite{bretherton1970note}. 

One may consider a particle relabeling $\vec{a}^\prime = \vec{a} + \delta\vec{a}(\vec{a},t)$ that does not change the particle paths, such that $\vec{x}^\prime(\vec{a}^\prime,t) = \vec{x}(\vec{a},t)$.
Under the relabeling, (\ref{eq:action}) must remain invariant, and 
in fact, we will further restrict the relabeling to those that do not alter the potential energy density $V$ itself.



\begin{figure}[h!]
    \centering
    \includegraphics[width=0.48\textwidth]{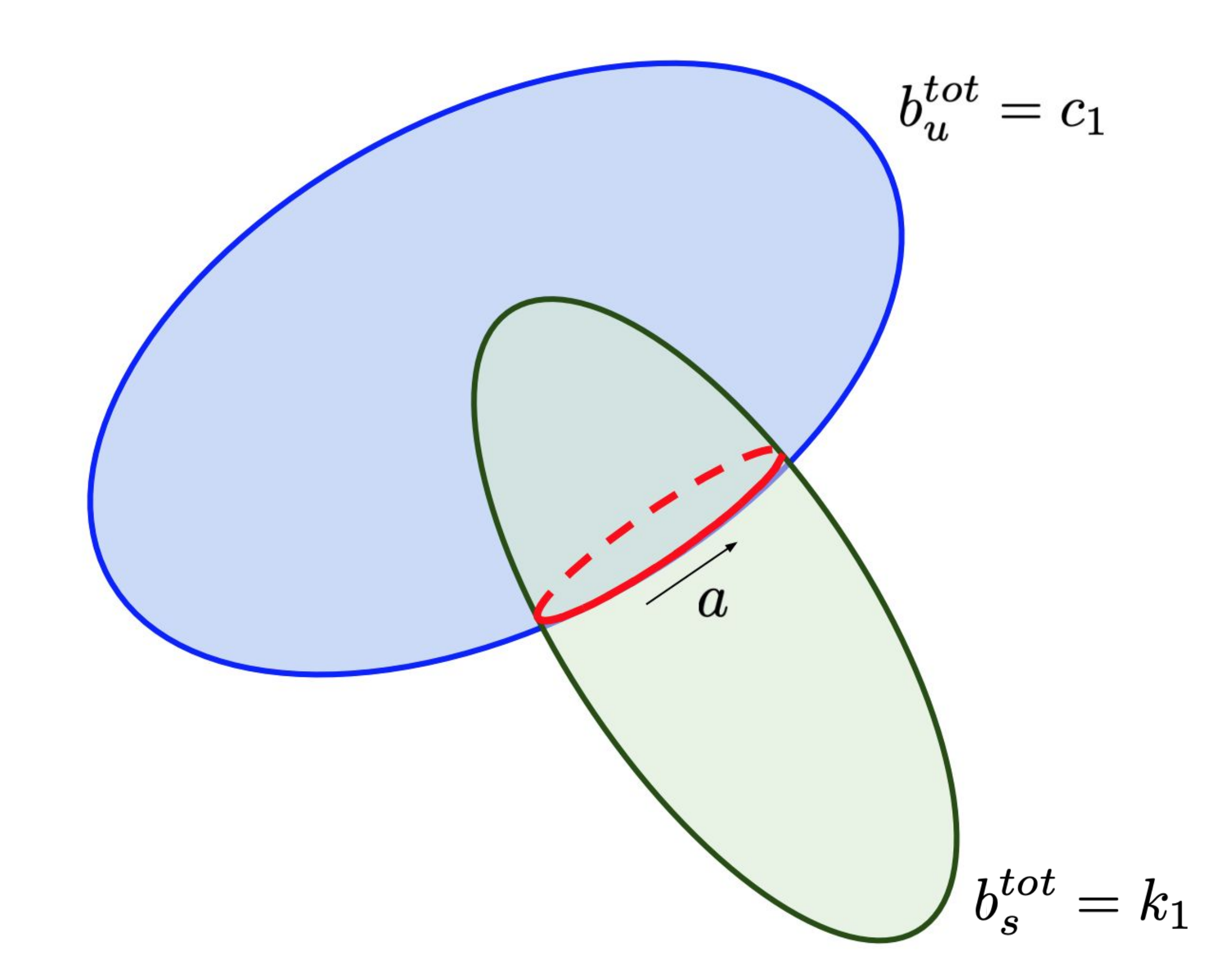}
    \includegraphics[width = 0.5\textwidth]{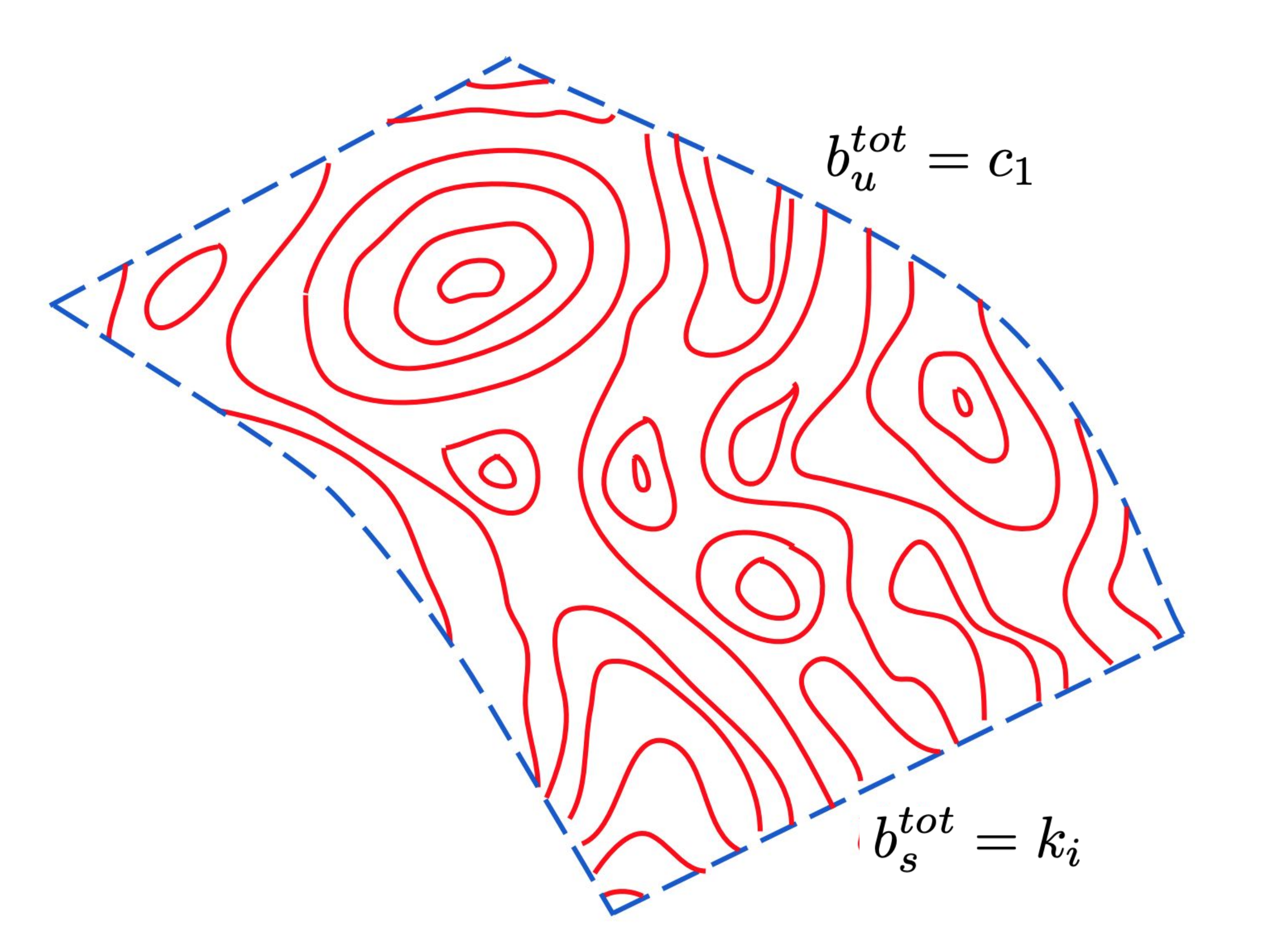}
    \caption{Left: Schematic diagram showing the particle labels, $\vec{a} = (a,b_u^{tot},b_s^{tot}).$
    Right: Schematic diagram showing level sets of $b_s^{tot}$ on a surface of constant $b_u^{tot}$, assuming that $b_s^{tot}$ and $b_u^{tot}$ are smooth functions.}
    \label{fig:labelcoordinates}
\end{figure}

To achieve invariance of potential energy, based on its form ${\cal V}(b_u^{tot},b_s^{tot},z)$, we see that we should consider variations $\delta\vec{a}$ along curves of constant $b_u^{tot}$ and constant $b_s^{tot}$, where a schematic of such curves is shown in figure \ref{fig:labelcoordinates}.  
These are curves at the intersection of two surfaces: one with constant $b_u^{tot}$ and the other with constant $b_s^{tot}$. 
Since surfaces of constant $b_u^{tot}$ and surfaces of 
constant $b_s^{tot}$ are both material invariant, it 
follows that their intersection is also material invariant. 
To simplify the variations of this form, it is convenient to consider the following label coordinate system: 
\begin{align}
    \vec{a} = (a,b,c) = (a, b_u^{tot}, b_s^{tot}).
\end{align}
 One can explicitly construct such curves with $\vec{a} = \vec{a}(\vec{x_0})$ depending on the initial particle label $\vec{x}_0$ associated with time $t=0$ (see Appendix A).
For a given particle, the first coordinate $a$ is defined to be an integral along a curve of constant $(b_u^{tot}, b_s^{tot})$,
starting from a reference point $\vec{x}_0^r$ on the curve, and ending at the initial location $\vec{x}_0$ of the particle at $t=0$. 
The label $a = a(\vec{x}_0)$ is chosen such that $dV_a =da\, db\, dc= da\, db_u^{tot}\, db_s^{tot}$ represents a mass element $\rho dV_x$, and furthermore satisfies the condition for material invariance, $Da/Dt=0$. 
Also note that $b$ had been used to represent buoyancy in some early parts of the paper, and $b$ is now being used in the label coordinate $\vec{a}=(a,b,c)$, and the meaning of $b$ should be clear from context in what follows.


Next, we must enforce the condition that the relabeling should not change the mass, such that
\begin{equation}
d\vec{a} = \frac{\partial (\vec{a})}{\partial (\vec{a}')}d\vec{a}^\prime = d\vec{a}^\prime,
\end{equation}
where 
\begin{equation}
    \frac{\partial (\vec{a})}{\partial (\vec{a}')} =  \frac{\partial (\vec{a}' - \delta \vec{a})}{\partial (\vec{a}')} 
    = 1 - \left(\frac{\partial\delta a}{\partial a} + \frac{\partial\delta b}{\partial b} + \frac{\partial\delta c}{\partial c}\right) + O(|\delta \vec{a}|^2) \quad {\rm for} \quad |\delta \vec{a}| \rightarrow 0.
    \label{eq:massjacobian}
\end{equation}
Neglecting the $O(|\delta \vec{a}|^2)$ truncation error leads to the condition
\begin{equation}
\label{eq:pvar2}
    \frac{\partial\delta a}{\partial a} + \frac{\partial\delta b}{\partial b} + \frac{\partial\delta c}{\partial c} = 0.
\end{equation}
By requiring (\ref{eq:pvar2}), together with $\delta\vec{a}$ occurring along curves of constant $(b_u^{tot}, b_s^{tot})$, leads to $\delta \vec{a}$ of the form
\begin{align}
\label{exp:relabeling}
    \delta {\vec{a}} = \left( \delta a (b,c,t), 0, 0  \right) = \left( \delta a (b_u^{tot},b_s^{tot},t), 0, 0  \right),
\end{align}
where $\delta a (b_u^{tot},b_s^{tot},t)$ is an arbitrary scalar function  with suitably small amplitude, and that vanishes at the boundaries of the time and space domain.
This form of the variation ensures a mass-conserving particle relabeling, up to an asymptotically small error in the limit 
$|\delta {\vec{a}}| \rightarrow 0.$
In brief, the relabeling shift $\delta a$ is a change in the label
coordinate $a$ of the location on the curve, and it is a uniform shift
for each curve because it is of the form
$\delta a (b_u^{tot},b_s^{tot},t)$ and does not depend on 
the label $a$ itself.

By adopting variations of the form (\ref{exp:relabeling}), we now proceed to enforce $\delta {\cal A}=0$, where $\delta {\cal A}$ arises from variations $\delta \vec{x}_t$ in the kinetic energy,
\begin{align}
\label{eq:deltaactionzero}
\delta \mathcal{A} &=  \int d t \iiint dV_a\;  \frac{\partial \vec{x}}{\partial t} \cdot \delta \frac{\partial \vec{x}}{\partial t}
= \int d t \iiint dV_a\;  \vec{x}_t \cdot \delta \vec{x}_t
= 0,
\end{align}
with the eventual aim of relating the result to conservation of potential vorticity.  
We remind the reader that $\vec{x}_t$ means the derivative of $\vec{x}(\vec{a},t)$ with respect to time $t$, keeping $\vec{a}$ fixed.  We will define $\delta \vec{x}_t$ below, and for clarity throughout the rest of this section, we will be explicit about what is held fixed when time derivatives are taken.
The steps in the rather lengthy calculation are (i) express $\delta\vec{x}_t$ in terms of 
$(\delta \vec{a})_t$; (ii) substitute expression (\ref{exp:relabeling}) for $\delta \vec{a}$; (iii) rearrange the ordering of terms in the integral; (iv) simplify the integral and interpret the result in physical terms.  As above in (\ref{eq:massjacobian}), it will be necessary to consider variations with $|\delta \vec{a}|\rightarrow 0$, and the final result is valid up to an asymptotically small error of $O(|\delta \vec{a}|)^2$.

An expression for $\delta\vec{x}_t$ in terms of $(\delta \vec{a})_t$ follows from calculus, followed by taking the limit $|\delta\vec{a}| \rightarrow 0.$ 
Starting from the definition of $\delta \vec{x}_t$, one may apply the chain rule:

\begin{align}
\delta \frac{\partial \vec{x}}{\partial t}&\equiv \frac{\partial  \vec{x}^\prime(\vec{a}^\prime,t) }{\partial t}\Bigr|_{\vec{a}^\prime} - \frac{\partial  \vec{x}(\vec{a},t) }{\partial t}\Bigr|_{\vec{a}} \\
&=\frac{\partial  \vec{x}(\vec{a},t)} {\partial t}\Bigr|_{\vec{a}^\prime} - \frac{\partial  \vec{x}(\vec{a},t) }{\partial t}\Bigr|_{\vec{a}} \\
&= \frac{\partial  \vec{x}(\vec{a},t)} {\partial t}\Bigr|_{\vec{a}} + \frac{\partial \vec{x}}{\partial a} \; \frac{\partial a}{\partial t}\Bigr|_{\vec{a}^\prime}  + \frac{\partial \vec{x}}{\partial b} \; \frac{\partial b}{\partial t}\Bigr|_{\vec{a}^\prime} + \frac{\partial \vec{x}}{\partial c} \; \frac{\partial c}{\partial t}\Bigr|_{\vec{a}^\prime}  - \frac{\partial  \vec{x}(\vec{a},t)} {\partial t}\Bigr|_{\vec{a}} \\
&= \frac{\partial \vec{x}}{\partial a} \; \frac{\partial a}{\partial t}\Bigr|_{\vec{a}^\prime}  + \frac{\partial \vec{x}}{\partial b} \; \frac{\partial b}{\partial t}\Bigr|_{\vec{a}^\prime} + \frac{\partial \vec{x}}{\partial c} \; \frac{\partial c}{\partial t}\Bigr|_{\vec{a}^\prime},
\label{eq:velocitydiff}
\end{align}

where $\nabla_{\vec{a}}  \vec{x} \; \partial \vec{a}/\partial t|_{\vec{a}^\prime}$ is a matrix-vector multiplication, 
and where we have used $\vec{x}^\prime(\vec{a}^\prime,t) = \vec{x}(\vec{a},t)$ and $\vec{a}^\prime = \vec{a} + \delta\vec{a}(\vec{a},t).$
Furthermore $\vec{a}^\prime = \vec{a} + \delta\vec{a}(\vec{a},t)$ implies that
\begin{align}
\label{eq:derivativeidentity}
\frac{\partial \vec{a}^\prime}{\partial t}\Bigr|_{\vec{a}^\prime} = 0 = 
\frac{\partial \vec{a}}{\partial t}\Bigr|_{\vec{a'}} +    \frac{\partial \delta \vec{a}}{\partial t}\Bigr|_{\vec{a'}}.
\end{align}
One may then re-arrange  (\ref{eq:derivativeidentity}) to find
\begin{align}
\frac{\partial \vec{a}}{\partial t}\Bigr|_{\vec{a}^\prime} =-\frac{\partial \delta \vec{a}}{\partial t}\Bigr|_{\vec{a}^\prime} \sim -    \frac{\partial \delta \vec{a}}{\partial t}\Bigr|_{\vec{a}} + O( |\delta \vec{a}|^2).
\label{eq:dadtfixedap}
\end{align}
Using (\ref{eq:dadtfixedap})
in (\ref{eq:velocitydiff}) leads to 
\begin{align}
\delta \frac{\partial \vec{x}}{\partial t}
&\sim -\left( \frac{\partial \vec{x}}{\partial a} \; \frac{\partial \delta a}{\partial t}\Bigr|_{\vec{a}}  + \frac{\partial \vec{x}}{\partial b} \; \frac{\partial \delta b}{\partial t}\Bigr|_{\vec{a}} + \frac{\partial \vec{x}}{\partial c} \; \frac{\partial \delta c}{\partial t}\Bigr|_{\vec{a}} \right) + O( |\delta \vec{a}|^2),
\end{align}
and for the special variations defined in (\ref{exp:relabeling}) we have,
\begin{align}
\delta \frac{\partial \vec{x}}{\partial t}
&\sim - \frac{\partial \vec{x}}{\partial a} \; \frac{\partial \delta a}{\partial t}\Bigr|_{\vec{a}} + O( |\delta a|^2).
\end{align}
Thus, asymptotically as $|\delta \vec{a}| \rightarrow 0$ and using integration by parts in $t$, the variation $\delta {\cal A}$ from the kinetic energy becomes
\begin{align}
\int d t \iiint dV_a\;  \frac{\partial \vec{x}}{\partial t} \cdot \delta \frac{\partial \vec{x}}{\partial t}
&\sim - \int d t \iiint dV_a\; 
\frac{\partial \vec{x}}{\partial t}\Bigr|_{\vec{a}} \cdot 
\biggl (\frac{\partial \vec{x}}{\partial a} \; \frac{\partial \delta a}{\partial t}\Bigr|_{\vec{a}} \biggr ) 
\label{(87)}\\
    &=
    - \int d t \iiint dV_a\; 
\frac{\partial \vec{x}}{\partial t} \cdot 
\frac{\partial \vec{x}}{\partial a} 
\frac{\partial}{\partial t} \delta a (b,c,t)
    \\
    &= 
    \int d t \iiint dV_a\; 
    \delta a (b,c,t)
    \frac{\partial}{\partial t}\left(
\frac{\partial \vec{x}}{\partial t}\cdot 
\frac{\partial \vec{x}}{\partial a} 
    \right)
    \\
    &= 
    \int d t \iint db\,dc\; 
    \delta a (b,c,t)
    \int da \;
    \frac{\partial}{\partial t}\left(
\frac{\partial \vec{x}}{\partial t}\cdot 
\frac{\partial \vec{x}}{\partial a} 
    \right),
\label{eq:deltaAkineticintermediate}
\end{align}
where the notation $|_{\vec{a}}$ was dropped after (\ref{(87)}) because all time
derivatives are now of the same type with $\vec{a}$ held fixed. 
Now recall that this variation (\ref{eq:deltaAkineticintermediate}) is set to zero, and  
since $\delta a(b,c,t)$ is arbitrary,
it follows that 
\begin{equation}
    \int da \;
    \frac{\partial}{\partial t}\left(
\frac{\partial \vec{x}}{\partial t}\cdot 
\frac{\partial \vec{x}}{\partial a} 
    \right)
    = 0.
\end{equation}
Note that $\partial/\partial t$ can be brought outside the
integral, since it is a time derivative with
the label coordinates $\vec{a}=(a,b,c)$ held fixed.
Also note that $\partial\vec{x}/\partial t$ is the velocity, $\vec{u}$, and 
consequently we arrive at
\begin{equation}
    \frac{d}{dt}\int 
\vec{u}\cdot 
\frac{\partial \vec{x}}{\partial a} \, da
    = 0.
\label{(92)}
\end{equation}
According to our definition of labels, $a$ parameterizes curves of constant $(b_u^{tot},b_s^{tot})$; see figure \ref{fig:labelcoordinates}. Therefore, returning from label space to $\vec{x}$-space and considering closed curves $C(b_u^{tot},b_s^{tot},t)$ of constant $(b_u^{tot},b_s^{tot})$, the result (\ref{(92)}) may be restated as
\begin{equation}
    \frac{D}{Dt}\oint_{C(b_u^{tot},b_s^{tot},t)}
\vec{u}\cdot d\vec{x}
    = 0,
\label{eqn:moistKelvinCirculationThm}
\end{equation}
where $d{\vec x}$ is the line element tangent to the curve $C(b_u^{tot},b_s^{tot},t)$. Statement (\ref{eqn:moistKelvinCirculationThm}) is the moist circulation theorem for the system (\ref{eqn:u-evol-bubs})-(\ref{eqn:incomp-bubs}) with phase changes.  It is the moist analogue of the Kelvin or Bjerknes circulation theorems from the dry case
\citep[e.g.,][]{mb02,cotter2014variational, thorpe2003bjerknes}.

Furthermore, it is possible to obtain conservation principles
for potential vorticity.
In this direction, using Stokes' theorem, we may rewrite the moist Kelvin circulation theorem (\ref{eqn:moistKelvinCirculationThm}) to obtain a conservation law for vorticity, $\vec{\omega}=\nabla\times\vec{u}$, on a material surface patch $S(\Vec{x}, t)$:
\begin{equation}
    \frac{D}{Dt} \iint_{S(\vec{x},t)} \Vec{\omega} \cdot d\Vec{S} = 0,
\end{equation}
where the boundary of the patch is the closed curve $C(b_u^{tot},b_s^{tot},t)$.
For the special case where $S(\Vec{x}, t)$ is on a surface of constant $b_u^{tot}$, we find
\begin{equation}
    \frac{D}{Dt} \iint_{S(\vec{x},t)} \Vec{\omega} \cdot \frac{\nabla b_u^{tot}}{\|\nabla b_u^{tot}\|} dS = 0.
\end{equation}
To then move toward a volumetric integral, take an additional 
integral with respect to $b_u^{tot}$ to yield
\begin{align}
    \int_{C_1}^{C_2}  \frac{D}{Dt} \iint_{S(\vec{x},t)} \Vec{\omega} \cdot \frac{\nabla b_u^{tot}}{\|\nabla b_u^{tot}\|} dS \; db_u^{tot} &= 0. 
    \label{eq:circulationafterStokes}
\end{align}
Since $b_u^{tot}$ is a label coordinate, and the material derivative $D/Dt$ is a time derivative holding labels fixed, we may rewrite (\ref{eq:circulationafterStokes}) as
\begin{equation}
    \frac{D}{Dt} \int \iint_{S(\vec{x},t)} \Vec{\omega} \cdot \frac{\nabla b_u^{tot}}{\|\nabla b_u^{tot}\|} dS \; db_u^{tot} =
    \frac{D}{Dt} \int \iint_{S(\vec{x},t)} \Vec{\omega} \cdot \frac{\nabla b_u^{tot}}{\|\nabla b_u^{tot}\|} dS \; \left(\|\nabla b_u^{tot}\| d\sigma \right) = 0,
\end{equation}
where $\sigma$ is arc length along curves in the direction of $\nabla b_u^{tot}$ (figure \ref{fig:pill_box}).
Simplifying, we find conservation of a volume-integrated moist potential vorticity $\int PV_u$ for a material volume which looks like a distorted cylinder or pill-box as in Figure \ref{fig:pill_box}:
\begin{equation}
    \frac{D}{Dt} \iiint_{V(\vec{x},t)} dV \; PV_u  = \frac{D}{Dt}\iiint_{V(\vec{x},t)} dV \; { \Vec{\omega} \cdot \nabla b_u^{tot}} = 0.
    \label{eq:patchintegratedPV-bu}
\end{equation}
In (\ref{eq:patchintegratedPV-bu}), the base and lid of the cylinder are surfaces of constant $b_u^{tot}$, connected by gradient lines of $b_u^{tot}$. Moreover, the cylinder can be viewed as a stack of patches, where each patch is on a different surface of constant $b_u^{tot}$, and the  boundary of each patch is a curve of constant $b_s^{tot}$. In this way, the $b_s^{tot}$ values on the sides of the cylinder can be written as functions $b_s^{tot}(b_u^{tot})$.
In the supplementary information, we also show that the conservation statement can be extended to a volume with sides given by any arbitrary function $b_s^{tot}(b_u^{tot})$.

Similarly, one can also arrive at conservation of volume-integrated moist potential vorticity $\int PV_s$:
\begin{equation}
  \frac{D}{Dt} \iiint_{V(\vec{x},t)} dV \; PV_s = \frac{D}{Dt} \iiint_{V(\vec{x},t)} dV \; { \Vec{\omega} \cdot \nabla b_s^{tot}} = 0,
    \label{eq:patchintegratedPV-bs}
\end{equation}
where the material volume in (\ref{eq:patchintegratedPV-bs}) has
base and lid which are surfaces of constant $b_s^{tot}$, connected by sides which are gradient lines of $b_s^{tot}$ and have $b_u^{tot} = b_u^{tot}(b_s^{tot})$.  

For the moist Boussinesq system (\ref{eqn:u-evol-bubs})-(\ref{eqn:incomp-bubs}), relations (\ref{eq:patchintegratedPV-bu}) and (\ref{eq:patchintegratedPV-bs}) are the strongest type of conservation statement for potential vorticity.  When phase changes are present, conservation applies following
local volumes of potential vorticity, enclosed by special surfaces defined in terms of $(b_u^{tot},b_s^{tot})$, rather than following individual fluid particles as in the dry system.


Now that the systematic procedure has produced the form of the 
circulation theorem and the conservation principle
for potential vorticity, it is possible to derive these results
directly from the moist Boussinesq evolution equations in
(\ref{eqn:u-evol-bubs})-(\ref{eqn:incomp-bubs}).
In the supporting information, 
we present direct derivations for the circulation theorem, globally-integrated $PV_u$, and parcel-integrated $PV_u$ and $PV_s$.
The direct derivations provide an independent confirmation
of these conservation principles.
The supporting information also includes special variations of the action leading to global conservation of $PV_u$ and $PV_s$.

\section{Discussion and Conclusions}
\label{sec:conclusions}

Here we have shown that Hamilton's Principle can be used to derive the moist Boussinesq dynamics (\ref{eqn:u-evol})--(\ref{eqn:incomp}), including phase changes between water vapor and liquid water. The Hamiltonian formulation is not obvious a priori because the buoyancy is piecewise, changing its functional form at interfaces between unsaturated and cloudy air. A key observation is that the piecewise potential energy associated with buoyancy is both continuous and differentiable.

After the Hamiltonian formulation was established, we were then able to investigate particle relabeling symmetry for the action associated with the moist dynamics, and thereby provide a systematic derivation of conservation statements for moist potential vorticity.   The conservation statements derived herein had not previously been found from other approaches.  Furthermore, the symmetry analysis naturally links the moist versions of Kelvin's circulation theorem and conservation of potential vorticity.

\begin{figure}[h!]
    \centering
    \includegraphics[width = 0.75\textwidth]{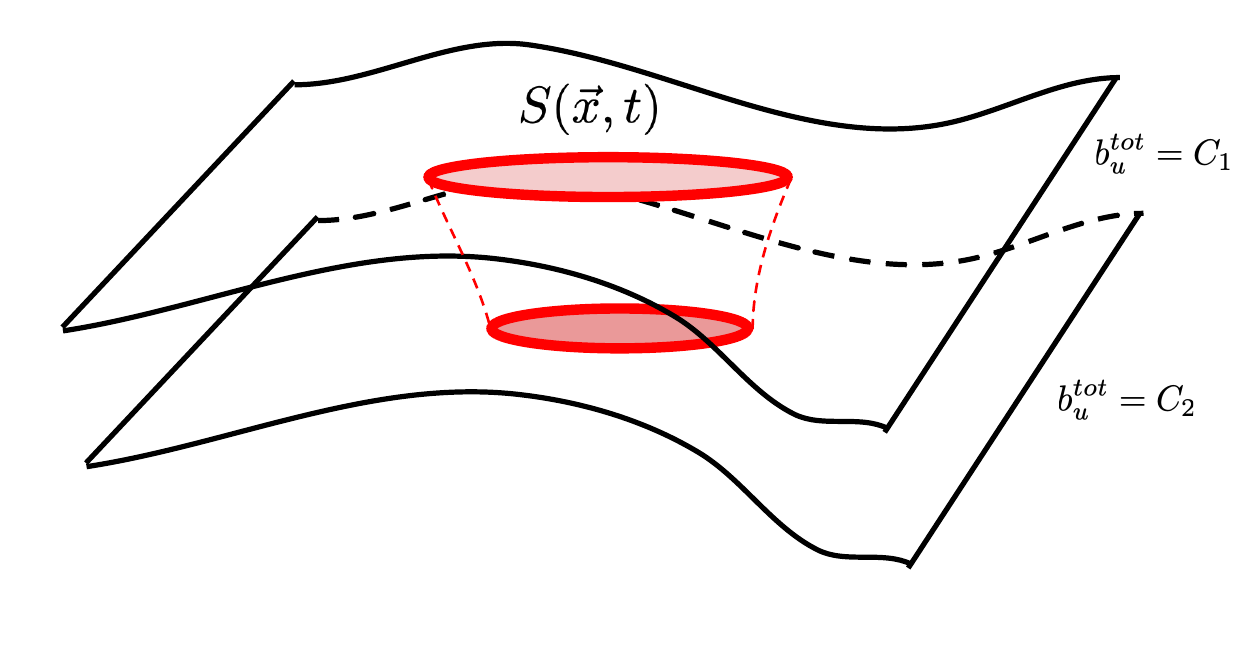}
    \caption{`Pancake'-like local material volume for parcel-integrated potential vorticity.}
    \label{fig:pill_box}
\end{figure}

Using particle relabeling symmetry, we uncover a fundamental difference between the dry and moist conservation statements for potential vorticity, following directly from the requirement that potential energy must remain invariant under the relabeling.  In the dry case,  it is necessary to restrict relabeling to constant entropy, leading to material invariance of dry potential vorticity on constant-entropy surfaces.
In the moist case with phase changes, the relabeling is restricted to fluid parcels situated on curves of constant $b_u^{tot}$ and constant $b_s^{tot}$. These curves are not restricted to the unsaturated or saturated regimes since $b_u$ and $b_s$ are globally defined quantities as in (\ref{eqn:bu-def}, \ref{eqn:bs-def}).
Ultimately, the stronger constraint leads to a weaker conservation law, namely, 
material invariance of parcel-integrated moist potential vorticity, where the local volumes have special surfaces defined in terms of $(b_u^{tot},b_s^{tot})$, as in Figure \ref{fig:pill_box}, for example.  Table 1 summarizes the dry and moist results, with respective definitions for potential vorticity. Note that one could restate the moist results in terms of any two material invariants for the system
(\ref{eqn:u-evol})--(\ref{eqn:incomp}), e.g., total potential temperature $\theta_e^{tot}$ and total water $q_t^{tot}$.

In previous work \citep[][]{marsico2019energy}, we established conservation of energy for the moist Boussinesq dynamics.  Here, we expand our knowledge of the relevant conservation laws to include a moist Kelvin circulation theorem and conservation of parcel-integrated moist potential vorticity, as well as their relationship to each other.  Thus, we now have moist analogues for some of the most important theorems for dry Boussinesq dynamics. 
Future work will explore how these results can be applied to analyze observations and numerical simulations of atmospheric flows.  It is straightforward to include the effects of rotation \citep[e.g.,][]{cotter2014variational}, which were omitted here for simplicity.  In addition, we have already generalized the statement for conservation of potential vorticity to the case of compressible moist flows, and these results will be presented elsewhere.  

\begin{table}[h!]
\caption{Summary of potential vorticity conservation principles for various regimes}
\begin{threeparttable}
\begin{tabular}{lll}
\headrow
\thead{Features} & \thead{Dry case (non-isentropic)} & \thead{Moist Boussinesq case} \\
Particle relabeling symmetry & On level surfaces $S_s(t)$ of entropy $s$. & On curves $C_b(t)$ of constant $b_u^{tot}$ and $b_s^{tot}$ \\
Circulation theorem:
$\oint_{C(t)} \vec{u} \cdot d\vec{x} = 0$ & $C(t):$ Any curve on $S_s(t)$&  $C(t) : $ Special curves $C_b(t)$ \\
PV definition & $\frac{1}{\rho} \vec{\omega} \cdot \nabla s$ & $PV_u=\vec{\omega} \cdot \nabla b_u^{tot}; PV_s=\vec{\omega} \cdot \nabla b_s^{tot}$ \\
Conserved quantity  & parcel PV & `pancake'-integrated PV 
\end{tabular}

\end{threeparttable}
\end{table}

\newpage
\section*{ACKNOWLEDGEMENTS}
Leslie Smith and Sam Stechmann gratefully acknowledge grant support from the Division of Mathematical Sciences of the National Science Foundation, in conjunction with the award DMS-1907667.

\section*{AUTHOR CONTRIBUTIONS}
All authors contributed to the problem formulation, mathematical calculations and preparation of the written manuscript.

\section*{CONFLICT OF INTEREST}
The authors declare no conflict of interest.

\section*{SUPPORTING INFORMATION}
In the supporting information, we include special variations of the action leading to global conservation of $PV_u$ and $PV_s$.
We also present direct derivations for the circulation theorem, globally-integrated $PV_u$, and parcel-integrated $PV_u$ and $PV_s$.

\vskip0.24in

\section*{Appendix A: Particle label definition}
\label{appendix-labeldefinition}

We can define the particle label $\vec{a}$ as follows:
\begin{align}
    \vec{a} = (a,b,c) = (a, b_u^{tot}, b_s^{tot}),
    \label{eq:newlabels}
\end{align}
where the first component $a$ should be such that it locates particles on curves of constant $(b_u^{tot}, b_s^{tot})$.
In addition, we require that
$dV_a = da\, db_u^{tot}\, db_s^{tot}$ represents a mass element in physical space with $dV_a =[\partial(\vec{a})/\partial (\vec{x})] dV_x= \rho dV_x$. 
Keeping in mind that the initial position vector
$\vec{x}_0$ is a mass conserving label with  $\partial(\vec{x_0})/\partial (\vec{x}) = \rho$, we 
therefore choose the first component $a$ in (\ref{eq:newlabels}) such that 
$\partial(\vec{a})/\partial (\vec{x}_0) = 1$.  A third requirement is for $a$ to be material invariant with $Da/Dt = 0$.
Below, we show an explicit construction for $\vec{a}$ satisfying all the above three restrictions.

\begin{figure}
    \centering
    \includegraphics[width=0.65\textwidth]{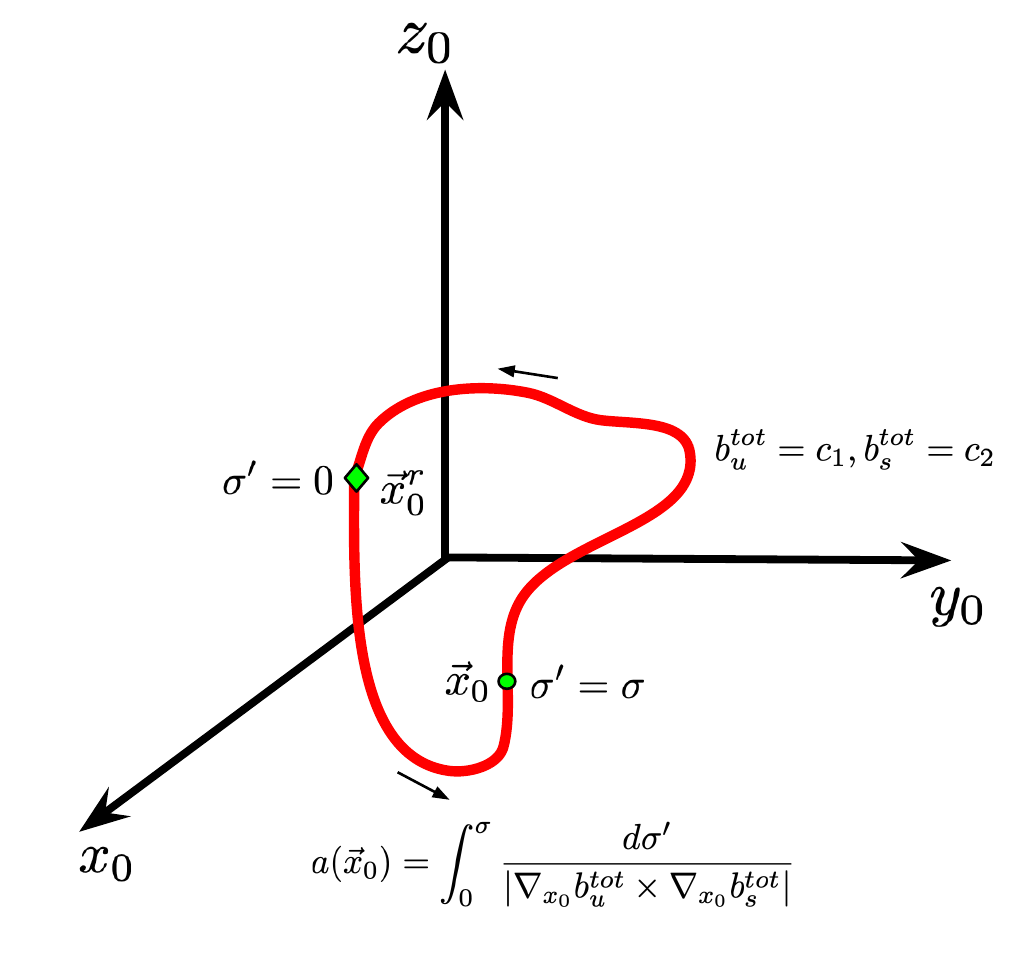}
    \caption{Schematic to illustrate the computation of the label $a(\vec{x}_0)$ in terms of initial position labels $\vec{x}_0$, where $a$ is the first coordinate of $\vec{a} = (a,b_u^{tot},b_s^{tot})$.}
    \label{fig:a_label}
\end{figure}

In order to concretely define the label $a$, start by drawing a parallelepiped in terms of $a, b_u^{tot}, b_s^{tot}$, where the direction of $a$ is determined by the cross product $\nabla_0 b_u^{tot} \times \nabla_0 b_s^{tot}$, and where $\nabla_0$ is a gradient with respect to the initial position vector $\vec{x}_0$ at time $t=0$. To enforce (\ref{def:determinantjacobian}), and using the chain rule, the magnitude of $a$ is determined by $\partial(\vec{a})/\partial (\vec{x}_0)=|\nabla_0 a \; \nabla_0 b_u^{tot}  \; \nabla_0 b_s^{tot}| = 1$ where $\nabla_0 a$ etc. are column vectors.   Therefore the label $a$ is chosen to satisfy the condition $|\nabla_0 a \; \nabla_0 b_u^{tot} \; \nabla_0 b_s^{tot}|=1$, which can be restated in terms of a scalar triple product as
\begin{equation}
(\nabla_0 b_u^{tot} \times \nabla_0 b_s^{tot}) \cdot \nabla_0 a = 1.
\label{eq:labelacondition}
\end{equation} 
The relation (\ref{eq:labelacondition}) will be true if we define
\begin{equation}
    \nabla_0 a = \frac{\nabla_0 b_u^{tot} \times \nabla_0 b_s^{tot}}{|\nabla_0 b_u^{tot} \times \nabla_0 b_s^{tot}|^2}. 
\end{equation}
Then integrating along a material curve $C$ of constant $(b_u^{tot}, b_s^{tot})$, the particle label $a$ is given by 
\begin{align}
    a(\vec{x}_0) &= \underset{C}{\int} \frac{\nabla_{x_0} b_u^{tot} \times \nabla_{x_0} b_s^{tot}}{|\nabla_{x_0} b_u^{tot} \times \nabla_{x_0} b_s^{tot}|^2} \cdot d\vec{x}_0' ,
\end{align}
where the integration is from reference point $\vec{x}_0^r$ to initial point $\vec{x}_0$.  Equivalently,  we may write
\begin{align}
    a(\vec{x}_0)&=\int_0^\sigma \frac{d\sigma'}{|\nabla_{x_0} b_u^{tot} \times \nabla_{x_0} b_s^{tot}|}, 
\end{align}
where $\sigma$ is the arc length along the curve $C$ (see figure \ref{fig:a_label}). Finally, notice that since $a$ is completely determined by $\vec{x}_0$, then $Da/Dt = 0$ by the chain rule, and therefore $a$ is material invariant as required.

\printendnotes
\bibliography{references,sambib}

\begin{thebibliography}{57}
\expandafter\ifx\csname natexlab\endcsname\relax\def\natexlab#1{#1}\fi
\expandafter\ifx\csname url\endcsname\relax
  \def\url#1{\texttt{#1}}\fi
\expandafter\ifx\csname urlprefix\endcsname\relax\def\urlprefix{URL: }\fi

\bibitem[{Abramov et~al.(2003)Abramov, Kova{\v{c}}i{\v{c}} and
  Majda}]{abramov2003hamiltonian}
Abramov, R.~V., Kova{\v{c}}i{\v{c}}, G. and Majda, A.~J. (2003) Hamiltonian
  structure and statistically relevant conserved quantities for the truncated
  {B}urgers-{H}opf equation.
\newblock \textit{Communications on Pure and Applied Mathematics: A Journal
  Issued by the Courant Institute of Mathematical Sciences}, \textbf{56},
  1--46.

\bibitem[{Bennetts and Hoskins(1979)}]{bennetts1979conditional}
Bennetts, D.~A. and Hoskins, B. (1979) Conditional symmetric instability-a
  possible explanation for frontal rainbands.
\newblock \textit{Quarterly Journal of the Royal Meteorological Society},
  \textbf{105}, 945--962.

\bibitem[{Brennan and Lackmann(2005)}]{brennan2005influence}
Brennan, M.~J. and Lackmann, G.~M. (2005) The influence of incipient latent
  heat release on the precipitation distribution of the 24--25 {J}anuary 2000
  {US} east coast cyclone.
\newblock \textit{Monthly Weather Review}, \textbf{133}, 1913--1937.

\bibitem[{Brennan et~al.(2008)Brennan, Lackmann and
  Mahoney}]{brennan2008potential}
Brennan, M.~J., Lackmann, G.~M. and Mahoney, K.~M. (2008) Potential vorticity
  {(PV)} thinking in operations: The utility of nonconservation.
\newblock \textit{Weather and Forecasting}, \textbf{23}, 168--182.

\bibitem[{Bretherton(1987)}]{b87i}
Bretherton, C.~S. (1987) {A theory for nonprecipitating moist convection
  between two parallel plates. Part I: Thermodynamics and "linear" solutions}.
\newblock \textit{Journal of the Atmospheric Sciences}, \textbf{44},
  1809--1827.

\bibitem[{Bretherton(1970)}]{bretherton1970note}
Bretherton, F.~P. (1970) A note on {H}amilton's principle for perfect fluids.
\newblock \textit{Journal of Fluid Mechanics}, \textbf{44}, 19--31.

\bibitem[{B{\"u}eler and Pfahl(2017)}]{bueler2017potential}
B{\"u}eler, D. and Pfahl, S. (2017) Potential vorticity diagnostics to quantify
  effects of latent heating in extratropical cyclones. {{P}art {I}:
  Methodology}.
\newblock \textit{Journal of the Atmospheric Sciences}, \textbf{74},
  3567--3590.

\bibitem[{Cao and Cho(1995)}]{cao1995generation}
Cao, Z. and Cho, H.-R. (1995) Generation of moist potential vorticity in
  extratropical cyclones.
\newblock \textit{Journal of the Atmospheric Sciences}, \textbf{52},
  3263--3282.

\bibitem[{Cotter(2013)}]{cotter2013data}
Cotter, C. (2013) Data assimilation on the exponentially accurate slow
  manifold.
\newblock \textit{Philosophical Transactions of the Royal Society A:
  Mathematical, Physical and Engineering Sciences}, \textbf{371}, 20120300.

\bibitem[{Cotter and Holm(2014)}]{cotter2014variational}
Cotter, C. and Holm, D. (2014) Variational formulations of sound-proof models.
\newblock \textit{Quarterly Journal of the Royal Meteorological Society},
  \textbf{140}, 1966--1973.

\bibitem[{Cuijpers and Duynkerke(1993)}]{cd93}
Cuijpers, J.~W.~M. and Duynkerke, P.~G. (1993) Large eddy simulation of trade
  wind cumulus clouds.
\newblock \textit{Journal of the Atmospheric Sciences}, \textbf{50},
  3894--3908.

\bibitem[{Davis and Emanuel(1991)}]{davis1991potential}
Davis, C.~A. and Emanuel, K.~A. (1991) Potential vorticity diagnostics of
  cyclogenesis.
\newblock \textit{Monthly Weather Review}, \textbf{119}, 1929--1953.

\bibitem[{Emanuel(1979)}]{emanuel1979inertial}
Emanuel, K.~A. (1979) Inertial instability and mesoscale convective systems.
  {P}art {I}: Linear theory of inertial instability in rotating viscous fluids.
\newblock \textit{Journal of the Atmospheric Sciences}, \textbf{36},
  2425--2449.

\bibitem[{Ertel(1942)}]{ertel1942neuer}
Ertel, H. (1942) Ein neuer hydrodynamischer wirbelsatz.
\newblock \textit{Meteorologische Zeitschrift}, \textbf{59}, 277--281.

\bibitem[{Gao et~al.(2004)Gao, Wang and Zhou}]{gao2004generation}
Gao, S., Wang, X. and Zhou, Y. (2004) Generation of generalized moist potential
  vorticity in a frictionless and moist adiabatic flow.
\newblock \textit{Geophysical Research Letters}, \textbf{31}.

\bibitem[{Hastermann et~al.(2021)Hastermann, Reinhardt, Klein and
  Reich}]{hastermann2021balanced}
Hastermann, G., Reinhardt, M., Klein, R. and Reich, S. (2021) Balanced data
  assimilation for highly oscillatory mechanical systems.
\newblock \textit{Communications in Applied Mathematics and Computational
  Science}, \textbf{16}, 119--154.

\bibitem[{Hernandez-Duenas et~al.(2013)Hernandez-Duenas, Majda, Smith and
  Stechmann}]{hmss13}
Hernandez-Duenas, G., Majda, A.~J., Smith, L.~M. and Stechmann, S.~N. (2013)
  Minimal models for precipitating turbulent convection.
\newblock \textit{J. Fluid Mech.}, \textbf{717}, 576--611.

\bibitem[{Hernandez-Duenas et~al.(2015)Hernandez-Duenas, Smith and
  Stechmann}]{hernandez2015stability}
Hernandez-Duenas, G., Smith, L.~M. and Stechmann, S.~N. (2015) Stability and
  instability criteria for idealized precipitating hydrodynamics.
\newblock \textit{Journal of the Atmospheric Sciences}, \textbf{72},
  2379--2393.

\bibitem[{Hill(1951)}]{hill1951hamilton}
Hill, E.~L. (1951) Hamilton's principle and the conservation theorems of
  mathematical physics.
\newblock \textit{Reviews of Modern Physics}, \textbf{23}, 253.

\bibitem[{Holm(2015)}]{holm2015variational}
Holm, D.~D. (2015) Variational principles for stochastic fluid dynamics.
\newblock \textit{Proceedings of the Royal Society A: Mathematical, Physical
  and Engineering Sciences}, \textbf{471}, 20140963.

\bibitem[{Holm et~al.(1998)Holm, Marsden and Ratiu}]{holm1998euler}
Holm, D.~D., Marsden, J.~E. and Ratiu, T.~S. (1998) The {E}uler--{P}oincar{\'e}
  equations and semidirect products with applications to continuum theories.
\newblock \textit{Advances in Mathematics}, \textbf{137}, 1--81.

\bibitem[{Hoskins and Berrisford(1988)}]{hoskins1988potential}
Hoskins, B. and Berrisford, P. (1988) A potential vorticity perspective of the
  storm of 15--16 {O}ctober 1987.
\newblock \textit{Weather}, \textbf{43}, 122--129.

\bibitem[{Hoskins(1991)}]{hoskins1991towards}
Hoskins, B.~J. (1991) Towards a {PV}-$\theta$ view of the general circulation.
\newblock \textit{Tellus A: Dynamic Meteorology and Oceanography}, \textbf{43},
  27--36.

\bibitem[{Hoskins et~al.(1985)Hoskins, McIntyre and Robertson}]{hoskins1985use}
Hoskins, B.~J., McIntyre, M.~E. and Robertson, A.~W. (1985) On the use and
  significance of isentropic potential vorticity maps.
\newblock \textit{Quarterly Journal of the Royal Meteorological Society},
  \textbf{111}, 877--946.

\bibitem[{Kuo(1961)}]{k61}
Kuo, H.~L. (1961) Convection in conditionally unstable atmosphere.
\newblock \textit{Tellus}, \textbf{13}, 441--459.

\bibitem[{Lackmann(2011)}]{lackmann2011midlatitude}
Lackmann, G. (2011) \textit{{M}idlatitude {S}ynoptic {M}eteorology: {D}ynamics,
  {A}nalysis, and {F}orecasting}.
\newblock American Meteorological Society.

\bibitem[{Lackmann(2002)}]{lackmann2002cold}
Lackmann, G.~M. (2002) Cold-frontal potential vorticity maxima, the low-level
  jet, and moisture transport in extratropical cyclones.
\newblock \textit{Monthly Weather Review}, \textbf{130}, 59--74.

\bibitem[{Madonna et~al.(2014)Madonna, Wernli, Joos and
  Martius}]{madonna2014warm}
Madonna, E., Wernli, H., Joos, H. and Martius, O. (2014) Warm conveyor belts in
  the era-interim dataset (1979--2010). {P}art {I}: Climatology and potential
  vorticity evolution.
\newblock \textit{Journal of Climate}, \textbf{27}, 3--26.

\bibitem[{Majda(2003)}]{m03}
Majda, A.~J. (2003) \textit{Introduction to {PDE}s and Waves for the Atmosphere
  and Ocean}, vol.~9 of \textit{Courant Lecture Notes in Mathematics}.
\newblock Providence: American Mathematical Society.

\bibitem[{Majda and Bertozzi(2002)}]{mb02}
Majda, A.~J. and Bertozzi, A.~L. (2002) \textit{Vorticity and Incompressible
  Flow}, vol.~27 of \textit{Cambridge Texts in Applied Mathematics}.
\newblock Cambridge: Cambridge University Press.

\bibitem[{Majda et~al.(2019)Majda, Moore and Qi}]{majda2019statistical}
Majda, A.~J., Moore, M. and Qi, D. (2019) Statistical dynamical model to
  predict extreme events and anomalous features in shallow water waves with
  abrupt depth change.
\newblock \textit{Proceedings of the National Academy of Sciences},
  \textbf{116}, 3982--3987.

\bibitem[{Marsico et~al.(2019)Marsico, Smith and Stechmann}]{marsico2019energy}
Marsico, D.~H., Smith, L.~M. and Stechmann, S.~N. (2019) Energy decompositions
  for moist {B}oussinesq and anelastic equations with phase changes.
\newblock \textit{Journal of the Atmospheric Sciences}, \textbf{76},
  3569--3587.

\bibitem[{Martin(2013)}]{martin2013mid}
Martin, J.~E. (2013) \textit{Mid-latitude {A}tmospheric {D}ynamics: {A} {F}irst
  {C}ourse}.
\newblock John Wiley \& Sons.

\bibitem[{Moore et~al.(2020)Moore, Bolles, Majda and Qi}]{moore2020anomalous}
Moore, M., Bolles, C.~T., Majda, A.~J. and Qi, D. (2020) Anomalous waves
  triggered by abrupt depth changes: {L}aboratory experiments and truncated
  {K}d{V} statistical mechanics.
\newblock \textit{arXiv preprint arXiv:2001.00995}.

\bibitem[{M{\"u}ller(1995)}]{muller1995ertel}
M{\"u}ller, P. (1995) Ertel's potential vorticity theorem in physical
  oceanography.
\newblock \textit{Reviews of Geophysics}, \textbf{33}, 67--97.

\bibitem[{Noether(1918)}]{emmy1918invariante}
Noether, E. (1918) Invariante variationsprobleme.
\newblock \textit{K{\"o}niglich Gesellschaft der Wissenschaften G{\"o}ttingen
  Nachrichten Mathematik-physik Klasse}, \textbf{2}, 235--267.

\bibitem[{Olver(2000)}]{olver2000applications}
Olver, P.~J. (2000) \textit{Applications of Lie {G}roups to {D}ifferential
  {E}quations}, vol. 107.
\newblock Springer Science \& Business Media.

\bibitem[{Padhye and Morrison(1996)}]{padhye1996relabeling}
Padhye, N. and Morrison, P. (1996) Relabeling symmetries in hydrodynamics and
  magnetohydrodynamics.
\newblock \textit{Tech. rep.}, Texas Univ., Austin, TX (United States). Inst.
  for Fusion Studies.

\bibitem[{Pauluis(2008)}]{p08}
Pauluis, O. (2008) Thermodynamic consistency of the anelastic approximation for
  a moist atmosphere.
\newblock \textit{J. Atmos. Sci.}, \textbf{65}, 2719--2729.

\bibitem[{Pauluis and Schumacher(2010)}]{ps10}
Pauluis, O. and Schumacher, J. (2010) Idealized moist {R}ayleigh--{B}{\'e}nard
  convection with piecewise linear equation of state.
\newblock \textit{Commun. Math. Sci}, \textbf{8}, 295--319.

\bibitem[{Pavlov et~al.(2011)Pavlov, Mullen, Tong, Kanso, Marsden and
  Desbrun}]{pavlov2011structure}
Pavlov, D., Mullen, P., Tong, Y., Kanso, E., Marsden, J.~E. and Desbrun, M.
  (2011) Structure-preserving discretization of incompressible fluids.
\newblock \textit{Physica D: Nonlinear Phenomena}, \textbf{240}, 443--458.

\bibitem[{Ripa(1981)}]{ripa1981symmetries}
Ripa, P. (1981) Symmetries and conservation laws for internal gravity waves.
\newblock In \textit{AIP Conference Proceedings}, vol.~76, 281--306. American
  Institute of Physics.

\bibitem[{Rossby(1939)}]{rossby1939relation}
Rossby, C.-G. (1939) Relation between variations in the intensity of the zonal
  circulation of the atmosphere and the displacements of the semi-permanent
  centers of action.
\newblock \textit{J. Mar. Res.}, \textbf{2}, 38--55.

\bibitem[{Salmon(1982)}]{salmon1982hamilton}
Salmon, R. (1982) Hamilton’s principle and {E}rtel’s theorem.
\newblock In \textit{AIP Conference Proceedings}, vol.~88, 127--135. American
  Institute of Physics.

\bibitem[{Salmon(1988)}]{salmon1988hamiltonian}
--- (1988) Hamiltonian fluid mechanics.
\newblock \textit{Annual {R}eview of {F}luid {M}echanics}, \textbf{20},
  225--256.

\bibitem[{Salmon(1998)}]{salmon1998lectures}
--- (1998) \textit{Lectures on {G}eophysical {F}luid {D}ynamics}.
\newblock Oxford University Press.

\bibitem[{Schubert et~al.(2001)Schubert, Hausman, Garcia, Ooyama and
  Kuo}]{schubert2001potential}
Schubert, W.~H., Hausman, S.~A., Garcia, M., Ooyama, K.~V. and Kuo, H.-C.
  (2001) Potential vorticity in a moist atmosphere.
\newblock \textit{Journal of the Atmospheric Sciences}, \textbf{58},
  3148--3157.

\bibitem[{Shepherd(1990)}]{shepherd1990symmetries}
Shepherd, T.~G. (1990) Symmetries, conservation laws, and {H}amiltonian
  structure in geophysical fluid dynamics.
\newblock In \textit{Advances in Geophysics}, vol.~32, 287--338. Elsevier.

\bibitem[{Smith and Stechmann(2017)}]{smith2017precipitating}
Smith, L.~M. and Stechmann, S.~N. (2017) Precipitating quasigeostrophic
  equations and potential vorticity inversion with phase changes.
\newblock \textit{Journal of the Atmospheric Sciences}, \textbf{74},
  3285--3303.

\bibitem[{Sommeria(1976)}]{s76}
Sommeria, G. (1976) Three-dimensional simulation of turbulent processes in an
  undisturbed trade wind boundary layer.
\newblock \textit{J. Atmos. Sci.}, \textbf{33}, 216--241.

\bibitem[{Spyksma et~al.(2006)Spyksma, Bartello and Yau}]{sby06}
Spyksma, K., Bartello, P. and Yau, M.~K. (2006) A {B}oussinesq moist turbulence
  model.
\newblock \textit{J. Turbulence}, \textbf{7}, 1--24.

\bibitem[{Thorpe(1985)}]{thorpe1985diagnosis}
Thorpe, A. (1985) Diagnosis of balanced vortex structure using potential
  vorticity.
\newblock \textit{Journal of the Atmospheric Sciences}, \textbf{42}, 397--406.

\bibitem[{Thorpe et~al.(2003)Thorpe, Volkert and
  Ziemia{\'n}ski}]{thorpe2003bjerknes}
Thorpe, A.~J., Volkert, H. and Ziemia{\'n}ski, M.~J. (2003) The {B}jerknes'
  circulation theorem: A historical perspective.
\newblock \textit{Bulletin of the American Meteorological Society},
  \textbf{84}, 471--480.

\bibitem[{Wetzel et~al.(2019)Wetzel, Smith, Stechmann and Martin}]{wssm19}
Wetzel, A.~N., Smith, L.~M., Stechmann, S.~N. and Martin, J.~E. (2019) Balanced
  and unbalanced components of moist atmospheric flows with phase changes.
\newblock \textit{Chin. Ann. Math. Ser. B}, \textbf{40}, 1005--1038.

\bibitem[{Wetzel et~al.(2020)Wetzel, Smith, Stechmann, Martin and
  Zhang}]{wetzel2020potential}
Wetzel, A.~N., Smith, L.~M., Stechmann, S.~N., Martin, J.~E. and Zhang, Y.
  (2020) Potential vorticity and balanced and unbalanced moisture.
\newblock \textit{Journal of the Atmospheric Sciences}, \textbf{77},
  1913--1931.

\bibitem[{Zhang et~al.(2021{\natexlab{a}})Zhang, Smith and
  Stechmann}]{zss21jfm}
Zhang, Y., Smith, L.~M. and Stechmann, S.~N. (2021{\natexlab{a}}) Effects of
  clouds and phase changes on fast-wave averaging: A numerical assessment.
\newblock \textit{Journal of Fluid Mechanics}, \textbf{920}, A49.

\bibitem[{Zhang et~al.(2021{\natexlab{b}})Zhang, Smith and
  Stechmann}]{zss21jnls}
--- (2021{\natexlab{b}}) Fast-wave averaging with phase changes: Asymptotics
  and application to moist atmospheric dynamics.
\newblock \textit{Journal of Nonlinear Science}, \textbf{31}, 38.

\end{thebibliography}

\end{document}